\documentclass[twocolumn,pra,aps,superscriptaddress,showpacs, flushbottom]{revtex4-1}
\usepackage{xspace,amsmath,amsfonts,amsthm,amssymb,amsbsy,bbm,graphicx,natbib}
\usepackage{booktabs}
\usepackage[usenames,dvipsnames]{color}
\usepackage{epstopdf}
\usepackage[colorlinks=true,
            urlcolor=black,
            citecolor=black,
            linkcolor=black]{hyperref}

\newcommand{\ident}{\boldsymbol{\mathbbm{1}}}

\newcommand{\ms}[1]{\mbox{\scriptsize #1}}
\newcommand{\msi}[1]{\mbox{\scriptsize\textit{#1}}}

\begin{document}
\title{Input-output theory for superconducting and photonic circuits that contain weak retro-reflections and other weak pseudo-cavities}

\author{Robert Cook}
\affiliation{U.S. Army Research Laboratory, Computational and Information Sciences Directorate, Adelphi, Maryland 20783, USA}
\author{David I. Schuster}
\affiliation{Department of Physics and James Franck Institute, University of Chicago, Chicago, Illinois 60637, USA}
\author{Andrew N. Cleland}
\affiliation{Institute for Molecular Engineering, University of Chicago, Chicago, Illinois 60637, USA}
\author{Kurt Jacobs}
\affiliation{U.S. Army Research Laboratory, Computational and Information Sciences Directorate, Adelphi, Maryland 20783, USA}
\affiliation{Department of Physics, University of Massachusetts at Boston, Boston, MA 02125, USA}
\affiliation{Hearne Institute for Theoretical Physics, Louisiana State University, Baton Rouge, LA 70803, USA}

\date{\today}

\begin{abstract}
 Input-output theory is invaluable for treating superconducting and photonic circuits connected by transmission lines or waveguides. However, this theory cannot in general handle situations in which retro-reflections from circuit components or configurations of beam-splitters create loops for the traveling-wave fields that connect the systems. Here, building upon the network-contraction theory of Gough and James [Commun. Math. Phys. \textbf{287}, 1109 (2009)], we provide a compact and powerful method to treat any circuit that contains such loops so long as the effective cavities formed by the loops are sufficiently weak. Essentially all present-day on-chip superconducting and photonic circuits will satisfy this weakness condition so long as the reflectors that form the loops are not especially highly reflecting. As an example we analyze the problem of transmitting entanglement between two qubits connected by a transmission line with imperfect circulators, a problem for which the new method is essential. We obtain a full solution for the optimal receiver given that the sender employs a simple turn on/turn off. This solution shows that near-perfect transmission is possible even with significant retro-reflections.
\end{abstract}

\pacs{03.67.-a, 42.50.Ex, 85.25.-j, 85.25.Cp}


\maketitle

\section{Introduction}

Input-output theory~\cite{Collett84, gardiner_input_1985, Holland90, gardiner_driving_1993, Kamal09, Lecocq17, Combes17, gardiner_quantum_2004, jacobs_quantum_2014}
is an important tool for describing the behavior of quantum superconducting~\cite{Kamal11, Naik17, Kandala17, Clark17, Fitzpatrick16, Kelly15} and photonic~\cite{Yoshie04, Reithmaier04, Fischer16, Schroder17} circuits. This theory models the interaction of circuit components (localized quantum systems) with transmission lines and wave guides that carry what are effectively traveling-wave fields. It provides a description in which the fields appear as ``inputs'' that drive the localized systems, and in which the fields that propagate away from the systems (the ``outputs'') contain both the input fields and a contribution from the systems. If the behavior of the systems is linear then the dynamics of a system or network of systems can be solved and the result is a single frequency-space ``scattering matrix'' that tells how the network transforms the input fields to the output fields as a function of frequency~\cite{Lecocq17, Holland90, Kamal09, Combes17, jacobs_quantum_2014}.

As useful as it is, input-output theory has an Achilles heel, in that it cannot in general handle situations in which the fields can traverse ``loops'' created inadvertently by retro-reflections from circuit components or deliberately through the use of beam-splitters. The reason that input-output theory breaks down in this situation is that such loops allow individual fields to interact repeatedly with the same system, potentially an infinite number of times, creating a ``non-Markovian'' dynamics in which states of the circuit components at one time are able to directly effect the components at later times via the fields~\cite{Grimsmo15,Pichler16}. Another way to understand this breakdown is that the effective cavities formed by the loops can change the mode structure of the fields so that they no-longer possess the simple continuum of modes on which input-output theory relies.

The method we present here begins with the observation that if the wave-packets emitted by the localized systems change slowly compared to the time that the input fields spend bouncing around within the network, then input-output theory should still provide a good description: on the timescale of the systems each system will see merely a single total field that is the sum over all the repeated traversals of the loops. Thus in the appropriate parameter regime, in which the ``ring-down" time of the fields within the network is sufficiently short, one should be able to obtain an effective input-output description for the circuit. It turns out that the mathematical machinery required to derive this description, with only a minor addition (the inclusion of inter-system phase shifts), is a ``network contraction'' theory already developed by Gough and James~\cite{gough_quantum_2009}. Here we extend and re-formulate this network contraction theory, as well as re-deriving it in the language of input-output theory familiar to physicists. The result is a simple and powerful tool for analyzing superconducting and photonic circuits that can handle internal reflections and other configurations in which the fields traverse loops (that is, traverse some arbitrarily complex network of interlocking effective weak cavities and ring cavities).

Gough and James developed a method to construct an effective, loop-free input-output description from a ``loopy'' network. This elegant mathematical theory is obtained by imposing the condition that the time delay in going from one system to another is zero. In developing their theory, Gough and James did not, however, establish the physical conditions, and thus the parameter regimes under which the dynamics of the resulting input-output network well-approximates that of the original network: the ring-down times depend not only on the travel-time between systems but also on the reflectivities that form the effective weak cavities. Our first main contribution is to derive these conditions in detail, showing at the level of the field commutators the regimes in which input-output theory can be expected to provide a valid description of a ``loopy'' network. In presenting their network contraction procedure, Gough and James emphasized its use in building a network one element at a time, which while appropriate for computer-automated computations is cumbersome if one wishes to perform calculations by hand. In extending (and specializing) the Gough-James method to superconducting and photonic circuits with weak retro-reflections and other weak loops, our second main contribution is to show explicitly how the entire network structure (the set of connections) can be captured by a single matrix. The effective input-output description of the network is then given by compact formulas in terms of this matrix. As the original derivation by Gough and James is not written in a language familiar to most physicists, we also both derive and present the method in such a language. 
We note that there is some overlap between the work presented here and concurrent work by Gough~\textit{et al.\ }presented in~\cite{gough_isolated_2017}.

In the second part of this paper we apply the method described above to the important problem of transferring a quantum state from one qubit to another via a uni-directional transmission line~\cite{wenner_catching_2014, Srinivasan14, yin_catch_2013, Bader13, korotkov_flying_2011, jahne_high-fidelity_2007, cirac_quantum_1997}. We show how this problem can be solved when we take into account that all the circuit elements, including the circulators that couple the qubits to the transmission lines, induce retro-reflections at their various interfaces. Such reflections cause unavoidable loss, the effect of which can be minimized by choosing the appropriate control protocol.

The rest of this paper is laid out as follows. In Sec.~\ref{sec:WeakCoupling} we discuss the derivation of the input-output formalism and in particular the approximations that it requires. These are important later when determining the conditions under which the method presented here is applicable. In Sec.~\ref{sec:IOnetwork} we explain how a set of unconnected input-output network elements is easily described using a single scattering matrix $\mathbf{S}$, a vector of operators $\mathbf{L}$, and a Hamiltonian $H$ (this formalism was introduced by Gough and James~\cite{gough_series_2009}). We follow this by explaining how to specify the connections between the elements (that is, define a network) using a single matrix $\mathbf{W}$. In Sec.~\ref{sec:effio} we show how to calculate the effective input-output description of a network, which is a single input-output system described by effective scattering matrix $\mathbf{S}_{\ms{eff}}$, vector of operators $\mathbf{L}_{\ms{eff}}$, and Hamiltonian $H_{\ms{eff}}$. We give the explicit expressions for these quantities in terms of $\mathbf{S}$, $\mathbf{L}$, $H$, and $\mathbf{W}$. We also describe the effective ``dissipative Hamiltonian'' for the effective input-output system, and we derive the regime in which the effective description is a good approximation. In Sec.~\ref{sec:example} we apply the method to the problem of transferring a quantum state from one qubit to another along a transmission line via imperfect circulators, something that is not possible with standard input-output theory. We show how the transmission probability can be maximized by controlling the receiver given that the coupling between the sender and the transmission line is simply turned on for a fixed amount of time. We also analyze the super- and sub-radiant states of this two-qubit network. In Sec.~\ref{sec:summary} we summarize our results and discuss open questions. The appendix gives the details of the algebraic manipulations required to derive the effective input-output model.







\begin{figure}[t]
    \begin{center}
    \includegraphics[width=0.95\hsize]{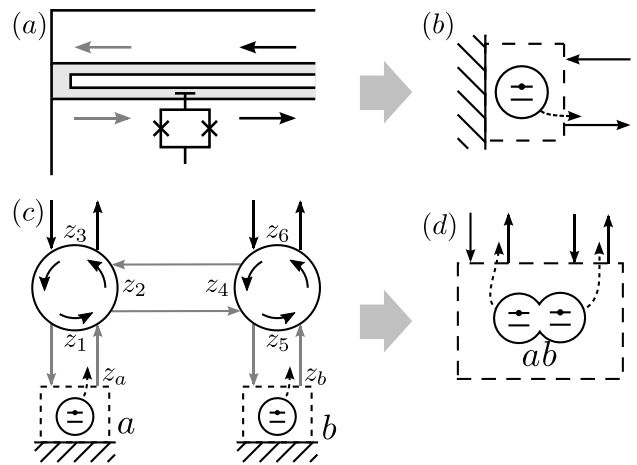}
        \caption{\label{fig:reductions} Two examples of superconducting circuits (networks) that contain weak loops, along with their corresponding reduced (loop-free) input-output (IO) models. The traveling-wave fields that are internal to the networks, and that are thus eliminated in obtaining the input-output descriptions, are denoted by solid grey arrows. The fields that are the external inputs and outputs to the networks are denoted by solid black arrows.  The outgoing emissions from the systems are depicted by dashed arrows. (a) A superconducting qubit capacitively coupled to a terminated planar waveguide, forming a quasi-1D input output system. (b) The effective IO model for (a) derived by eliminating the internal field modes (grey arrows). (c) Two of the qubit IO models in (b) connected together via imperfect circulators. (d) The reduced two-qubit IO model for (c) obtained by eliminating all the internal fields.}
	\end{center}
\end{figure}

\section{Input-output theory: local systems weakly interacting with traveling-wave fields} \label{sec:WeakCoupling}

Here we discuss the key assumptions and approximations that lead to the equations of input-output theory~\cite{gardiner_quantum_2004, jacobs_quantum_2014}. We will need to refer to these assumptions when we derive the conditions under which a network with internal reflections can be well-approximated by an effective input-output description. Since the work here builds upon input-output theory, some basic familiarity with this theory will be helpful to the reader. In particular we recall that input-output theory describes a localized system interacting with a 1D, unidirectional propagating field. This theory provides two key equations, one which gives the output field in terms of the input field and the emissions from the system, and a second that gives the dynamics of the system as a Heisenberg-Langevin equation driven by the field. For references purposes we note that the input-output relation for the field is given in Eq.(\ref{eq:1DIO}) (for an interaction with a single field) and Eq.(\ref{eq:genio}) (for a set of interactions with multiple 1D fields). The Heisenberg-Langevin equation for one or more systems interacting with the fields is given in Eq.(\ref{eq:HLeqs}).

Input-output theory is also able to handle a ``network'' situation in which the output field from one system is connected to the input field of another system. This was first shown by Gardiner~\cite{gardiner_driving_1993}, who referred to such a configuration as ``cascading'' two quantum systems together. Here we will be considering a lossless network of guided 1D fields that connect local quantum systems in this manner.


We will work with a general 1D scalar field, $F(z)$, which when written in terms of the right ($+$) and left ($-$) propagating modes, is
\begin{equation}\label{eq:Field}
\begin{split}
 F(z) &=  \int_{0}^\infty \frac{d \omega}{\sqrt{2 \pi}} \mathcal{F}(\omega) \left[ b_{+}(\omega) e^{i \omega z /v_p} + b_{-}(\omega) e^{-i \omega z/v_p} \right]  \\
  &\quad + h.c.,
\end{split}
\end{equation}
where $\mathcal{F}(\omega)$ is the complex vacuum field strength, $v_p$ is the phase velocity, and $b_{\pm}(\omega)$ are canonical commuting field operators with units of $1/\sqrt{\text{Hz}}$.

We wish to consider quantum systems that couple to this field via a linear interaction of the form~\cite{gardiner_quantum_2004}
\begin{equation}\label{eq:Hsf}
  H_{SF} = - Q_i F(z_i),
\end{equation}
for a system operator $Q_i$ located at position $z_i$.  Our model assumes that the system contains only a single dominant resonance frequency $\omega_0$, which for simplicity, corresponds to the energy gap between the relevant system states $|g\rangle$ and $|e\rangle$.  Furthermore, $Q_i$ is assumed to only contains off-diagonal matrix elements coupling $|g\rangle$ and $|e\rangle$.  Note that the resonant interaction strength, $\langle e|Q_i|g\rangle \mathcal{F}(\omega_0)/\hbar $ has units of $\sqrt{\text{Hz}}$.  The frequency $\kappa_0 \equiv |\langle e|Q_i|g\rangle \mathcal{F}(\omega_0)|^2/\hbar^2$ is ultimately the rate at which $|e\rangle$ decays into $|g\rangle$ by emitting an excitation into the field.  After transforming to the interaction picture and dropping any counter rotating terms, the interaction Hamiltonian is
\begin{equation}
\begin{split}
  H_{SF} &= - i \hbar \sqrt{\kappa_0} \sigma^{+}_{i} e^{i \phi} \left[ e^{i k_0 z_i } b_{+}(z_i, t) + e^{-i k_0 z_i} b_{-}(z_i, t) \right]\\
  &\quad + h.c.,
\end{split}
\end{equation}
where $\sigma^{+}_{i} = |e\rangle \langle g|$ and $\phi$ is the phase angle that sets the coupling quadrature, i.e.\ $\phi = \arg\, \langle e|Q_i|g\rangle \mathcal{F}(\omega_0)/ i \hbar$.  The field operators $b_{\pm}(z_i, t)$ are defined as,
\begin{equation}\label{eq:bofz}
  b_{\pm}(z_i, t) \equiv \int_{0}^\infty \frac{d\omega}{\sqrt{2 \pi}} \frac{\mathcal{F}(\omega)}{\mathcal{F}(\omega_0)} b_{\pm}(\omega) e^{- i (\omega - \omega_0)(t \mp z_i/v_p)}.
\end{equation}
In this definition we have factored out the carrier traveling waves $e^{\pm i k_0 z - i \omega_0 t}$ (with wave number $k_0 = \omega_0/v_p$).  However, in moving to the interaction picture, $\sigma^{+}_i$ generated the phase $e^{i \omega_0 t}$, canceling the time dependence of the carriers.

Standard input output theory, considers only a single system localized at a given position and thus any spatial phases can be safely ignored. However when considering two systems coupling to the same field at differing positions, propagation phases become relevant. The nonzero commutation relations between the field operators at unequal positions and times are
\begin{multline}
  [b_{\pm}(z_i, t), b^\dag_{\pm}(z_j, t')] =  \int_{0}^\infty \frac{d\omega}{2 \pi}\frac{|\mathcal{F}(\omega)|^2}{|\mathcal{F}(\omega_0)|^2} \\
   \times \exp \left[ -i (\omega - \omega_0)( t - t' \mp (z_i - z_j)/v_p ) \right].
\end{multline}
(The counter propagating fields commute with $[b_{\pm}(z_i, t), b^\dag_{\mp}(z_j, t')] = 0$, as they integrate over wavevectors with opposite signs.)

Here we will make the crucial approximation that this commutation relation will ultimately be approximated by a delta function in time.  Whether or not this approximation is made with respect to the absolute time, or a retarded time ultimately resides in the relevant time and distance scales.  We have already assumed that the relevant time scale is given by $1/\kappa_0$ thus we define the unitless time and frequency variables
\begin{equation}
  \Delta \tau \equiv (t - t') \kappa_0 \qquad \nu \equiv (\omega - \omega_0)/ \kappa_0,
\end{equation}
which are both $\sim O(1)$.  After making this change of variables and using the relation that $v_p = \omega_0/k_0$,
\begin{multline}
  [b_{\pm}(z_i, t), b^\dag_{\pm}(z_j, t')] =  \int_{-\tfrac{\omega_0}{\kappa_0}}^\infty \frac{d \nu}{2 \pi} \frac{\kappa_0}{|\mathcal{F}(\omega_0)|^2} \\
   \times \left|\mathcal{F}\left(\omega_0( 1 +  \nu \tfrac{\kappa_0}{\omega_0}) \right)\right|^2 e^{-i \nu ( \Delta \tau  \mp (z_i - z_j)k_0 \kappa_0/\omega_0) }.
\end{multline}
The assumption of weak coupling implies that $\kappa_0 \ll \omega_0$.  The spatial dependence of this commutator comes down to the comparison of the separation $|z_i - z_j|$ to the coherence length $\ell_0 \equiv v_p/\kappa_0$.  Here we focus on the case where $|z_i - z_j| \ll \ell_0$, or equivalently, $|z_i - z_j|k_0 \ll \omega_0/\kappa_0$.  Note that this is a significantly weaker criteria than is usually imposed for either a lumped element circuit model or free space superradiance~\cite{dicke_coherence_1954}, which both assume that the distance between systems is small when compared to the wavelength, not merely this larger coherence length.

Current experiments with superconducting qubits~\cite{wenner_catching_2014, Axline18} operate at $\omega_0 \sim 2 \pi \times 6 \text{ GHz}$ with maximum coupling rates varying between $\kappa_0 \simeq 2 \pi \times 100 \text{ MHz}$~\cite{wenner_catching_2014} and $\kappa_0 \simeq 2 \pi \times 400 \text{ KHz}$~\cite{Axline18}.  These experiments generally satisfy the criteria of weak coupling as $10^{-6} \lesssim \kappa_0/\omega_0 \lesssim 10^{-2}$.  In terms of the coherence length $ 2 \text{ m} \lesssim \ell_0 \lesssim 500 \text{ m}$, given a typical phase velocity $v_p = 0.7 c$.

In the limit $\kappa_0/\omega_0 \rightarrow 0$, the approximation
\begin{equation}
\begin{split}
  [b_{\pm}(z_i, t), b^\dag_{\pm}(z_j, t')] & \approx \kappa_0 \int_{- \infty}^\infty \frac{d \nu}{2 \pi} e^{-i \nu \Delta \tau } = \kappa_0 \, \delta(\Delta \tau ) \\
    & = \delta(t - t'),
\end{split}
\end{equation}
becomes exact. In light of this, we will approximate the spatially dependent field operators by a single one, $b_{\pm}(z_i, t) \approx b_{\pm}(t)$, resulting in a monochromatic approximation for the free field;
\begin{equation}\label{eq:MonoField}
\begin{split}
  F(z, t) &\approx \mathcal{F}(\omega_0) e^{i k_0 z - i \omega_0 t} b_{+}(t) + h.c. \\
    &\ + \mathcal{F}(\omega_0) e^{-i k_0 z - i \omega_0 t} b_{-}(t) + h.c.\\
    & \equiv F_{+}(z,t) + F_{-}(z,t)
\end{split}
\end{equation}

For an ensemble of $N$ systems identically coupled to the same waveguide, this approximation results in the total interaction Hamiltonian
\begin{equation}\label{eq:HsfTot}
  H_{SF} = - i \hbar \left[  L_{+}^\dag  b_{+}(t) + L_{-}^\dag  b_{-}(t) \right] + h.c.,
\end{equation}
where the collective excitation operators are:
\begin{equation}\label{eq:Lpm}
  L_{\pm}^\dag = \sqrt{\kappa_0} e^{i \phi} \sum_{i = 1}^{N} \sigma^{+}_{i} e^{\pm i k_0 z_i}.
\end{equation}
We have now covered the assumptions of input-output theory to the extent we need for our analysis below. To obtain the equations of input-output theory from this point onwards we refer the reader to the standard treatment (see, for example~\cite{gardiner_quantum_2004,jacobs_quantum_2014}).

\section{Defining a network of input-output systems} \label{sec:IOnetwork}

We now show how one can specify a network of input-output systems (that is, specify how the inputs and outputs of the systems are connected together) using a single matrix. To do this one first decides on a set of input-output systems (``network elements") that are to be connected together to form the network. These may be systems with their own internal dynamics that are coupled to transmission lines or wave guides, or they may be beam-splitters that merely connect various wave-guides or transmission lines together. These latter elements essentially set boundary conditions that the traveling-wave fields must obey. As we will explain, a given set of network elements can be described by i) a ``scattering matrix'', ii) a vector whose elements are the operators by which the dynamical systems are coupled to the fields, and iii) the Hamiltonians of the systems. Given this compact description of the network elements, the connections between the inputs and outputs of the elements, which completely define the network, are then captured by a single matrix.

\subsection{The 1D input-output relation} \label{ss:1Dio}
For a system coupled to a single IO channel, e.g.\ a superconducting qubit terminating a 1D transmission line, the detection of an outgoing photon could have one of two possible origins. Either the qubit system made a transition creating an outgoing photon, or an incoming photon reflected off the terminating boundary condition.  In a lossless system, a perfectly reflecting boundary can only result in a scattering phase shift between the incident and exiting fields.  The Heisenberg equation of motion for the outgoing field operator will be the coherent sum of these two processes, i.e.\
\begin{equation}\label{eq:1DIO}
  b_{out}(t) = e^{i \theta} b_{in}(t) + L(t).
\end{equation}
Here $\theta$ is the scattering phase shift and $L(t)$ is a system operator describing resonant emission.

As an example consider Fig.~\ref{fig:reductions}$(a)$.  In order for an interaction Hamiltonian like Eq.(\ref{eq:Hsf}) to implement a single channel model, the macroscopic boundary conditions that describe perfect reflection at the position $z_0$, require that $F(z) = 0$ for all $z \le z_0$.  Setting $F(z_0)$ in Eq.(\ref{eq:MonoField}) to zero leads to the boundary condition
\begin{equation}
  b_{+}(t) = - e^{-i 2 k_0 z_0} b_{-}(t),
\end{equation}
or in other words $\theta = -2 k_0 z_0 + \pi$.  This boundary has a physical effect on system's emission, as the reflected portion of the left going part interferes with the right going part. In fact direct substitution into Eq.(\ref{eq:HsfTot}), with $N = 1$, results in
\begin{equation}\label{eq:H1D}
  H_{1D} = - i \hbar \left[ L^\dag_{1D} b_{-}(t) -  L_{1D} b^\dag_{-}(t) \right]
\end{equation}
where
\begin{equation}
  L_{1D} = 2 \sqrt{\kappa_0} e^{- i \phi - i \theta/2} \cos[ k_0 z_i + \theta/2 ]\, \sigma^{-}.
\end{equation}
Thus a change in $\theta$ has a real effect on the system field coupling, possibly transforming a system located at an anti-node to a node where $L_{1D} = 0$. The importance of the above example is that we can utilize the lesson of the reflecting boundary condition at $z_0$ to apply multiple constraints across a network of guided modes.

\subsection{The Scattering Matrix}  \label{ss:S}
Utilizing the lessons of microwave engineering, the linear passive response of a lossless multiport device is summarized by a single unitary scattering matrix $\mathbf{S}$, which maps the incoming traveling waves to the outgoing ones.  Here we adopt the convention that the outgoing traveling wave exiting a port will have the same mode index as the field entering that port.  Thus when $\mathbf{S}_{ij} = \delta_{ij}$ all incoming fields are perfectly reflected with no change of phase ($\theta_i = 0$).  This is in contrast to a convention where $\delta_{ij}$ corresponds to perfect transmission in some preferred direction.   However, our results are ultimately independent from any particular choice of mode labels.  

A pertinent example for a multiple port scatter is an imperfect circulator.  Fig.\ref{fig:reductions}(c) shows two localized qubit systems connected via two circulators. Consider the circulator on the left, for which the spatial coordinates of the three ports are denoted by $z_1$, $z_2$, and $z_3$, with $z_1$ and $z_3$ lying on a vertical axis increasing from bottom to top and $z_2$ on a horizontal axis increase from left to right.  If we collect the input and output fields of the three ports into vectors as 
\begin{align}
	\mathbf{b}_{\ms{in}}(t) & = \left(\begin{array}{c}
        							b_+(z_1, t) \\
        							b_-(z_2, t) \\
        							b_-(z_3, t) \\
   							 \end{array}\right)\! , \;\; 
   \mathbf{b}_{\ms{out}}(t) = \left(\begin{array}{c}
        b_-(z_1, t) \\ 
        b_+(z_2, t) \\
        b_+(z_3, t) \\
   \end{array}\right) \! , 
\end{align}
then the scattering matrix tells us how the input fields get split up and directed among the output fields:  
\begin{equation}
  \mathbf{b}_{\ms{out}}(t) = \mathbf{S}_{\ms{circ}}\, \mathbf{b}_{\ms{in}}(t) .
\end{equation}
Given how we have chosen to order the input and output fields within the vectors $\mathbf{b}_{\ms{in}}(t)$ and $\mathbf{b}_{\ms{out}}(t)$, if we write the scattering matrix $\mathbf{S}_{\ms{circ}}$ as 
\begin{equation}
  \mathbf{S}_{\ms{circ}} =  \left( 
                             \begin{array}{ccc}
                               r_{11} & c_{12} & t_{13} \\
                               t_{21} & r_{22} & c_{23} \\
                               c_{31} & t_{32} & r_{33} \\
                             \end{array}
                           \right) 
\end{equation}
then the elements $r_{ii}$ are the input retroreflections, the $t_{ij}$ are the circulating transmission coefficients, and $c_{jk}$ are the ``cross-talk'' coefficients.  

A crucial point for the formalism we develop is that we only consider scattering matrices that describe classical boundary conditions and are independent from any quantum degrees of freedom.  In other words, we assume that for any quantum system operator $A$, $[\mathbf{S}, A] = 0$. The original theory of Gough and James is more general in that they used the quantum probability theory of Hudson and Parthasarathy~\cite{hudson_quantum_1984} to include theoretical models in which $[\mathbf{S}, A] \not= 0$.

\subsection{The many-input many-output relation} \label{ss:NDio}
\label{mimo}

The many-input analogy of the single-field IO relation, Eq.(\ref{eq:1DIO}), is made by first defining, for an $N$ port system, the vector of inputs (outputs), 
\begin{equation}
  \mathbf{b}_{\ms{in}}(t) =
\left(  \begin{array}{c}
b_{\lambda_1}(z_1, t)  \\
b_{\lambda_2}(z_2, t) \\
\vdots \\
b_{\lambda_N}(z_N, t)
\end{array}
\right) \! , \;\;
  \mathbf{b}_{\ms{out}}(t) =
 \left(  \begin{array}{c}
b_{\bar\lambda_1}(z_1, t)  \\
b_{\bar\lambda_2}(z_2, t) \\
\vdots \\
b_{\bar\lambda_N}(z_N, t)
\end{array}
\right)  \!, 
\end{equation}
where the propagation direction $\lambda_i$ is positive (negative) when the coordinate $z_i$ is increasing (decreasing) as the input field approaches the port.  The corresponding output field then propagates in the opposite direction, thus $\bar\lambda_i \equiv -\lambda_i$.   The general network IO relation is then
\begin{equation}
  \mathbf{b}_{\ms{out}}(t) = \mathbf{S}\mathbf{b}_{\ms{in}}(t) + \mathbf{L}(t) .
  \label{eq:genio}
\end{equation}
Here $\mathbf{L}$ is a vector composed of the operators via which the systems interact with the fields at each of their ports:
\begin{equation}
  \mathbf{L}(t) =  \left(  \begin{array}{c}
                 L_1(t) \\
                 L_2(t)  \\
                  \vdots \\
                  L_N(t)
                  \end{array} \right) .
\end{equation}
Each of the operators $L_i$ describes the resonant emission from a quantum system inside the scattering region.  Note that the elements of $\mathbf{L}$ are labeled, not by the individual subsystems, but by the \emph{output} ports.  If our model contains multiple subsystems within the scattering region, then the $L_i(t)$ will in general be collective operators (e.g.\ $L_{\pm}$ in Eq.(\ref{eq:Lpm})). 

\begin{figure}[tbp]
    \begin{center}
    \includegraphics[width=0.95\hsize]{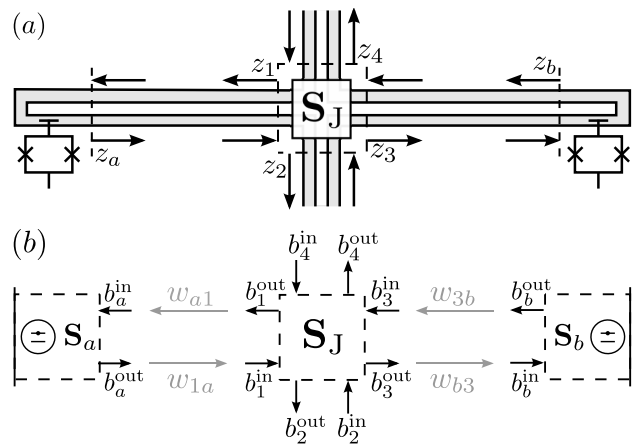}
        \caption{\label{fig:cross} A two-qubit ``crossed waveguide'' network.  (a) A schematic depicting two superconducting circuits weakly coupled to a pair of crossed waveguides, with a general unitary scattering relation.  (b) A block diagram of all field operators.  Input field operators $b^{\ms{in}}_{j}$ are scattered to the output fields $b^{\ms{out}}_{k}$ via the local elements of $\mathbf{S}$, e.g. $\mathbf{S}_{J}$, and irrespective of the waveguides.  The internal inputs are connected to the internal outputs by the connection matrix elements $w_{kl}$ (gray).}
	\end{center}
\end{figure}

Fig.~\ref{fig:cross} shows an example multimode network containing two crossed waveguides that form a general 4-input/4-output junction.  The local scattering between inputs $b^{\ms{in}}_{i}$ and $b^{\ms{out}}_{j}$, $i,j = 1 \dots 4$, is made by a general 4-by-4 unitary matrix $\mathbf{S}_{\ms{J}}$. The network also contains two superconducting qubits that are each coupled to one of the ports of the 4-port junction. To describe this network we first consider the two qubits and the 4-port junction as three separate components, each with their own input/output ports. Since each of the qubits has a single input/output port, together the three components have 6 inputs and 6 outputs.

We now define a scattering matrix $\mathbf{S}$ that describes the relationship between the 6 inputs and outputs \textit{before} the three components are connected together to form the network. Since each of the components acts on its own inputs separately from the others, this scattering matrix is block-diagonal where each block describes the action of one of the components. Since each qubit has only one port, $\mathbf{S}$ has two 1-by-1 blocks and a single 4-by-4 block given by $\mathbf{S}_{\ms{J}}$:
\begin{equation}
\mathbf{S} = \left(
               \begin{array}{c|c}
                 \begin{array}{cc} e^{i\theta_a} & 0  \\ 0 & e^{i\theta_b}  \\  \end{array}  & \scalebox{1.4}{ $\mathbf{0}$ } \!\! \\
                  \hline
                  \raisebox{1.2ex}[2.6\height][0\depth]{\scalebox{1.4}{ $\mathbf{0}$  }}  & \raisebox{1.25ex}[2.6\height][0\depth]{\scalebox{1.25}{  $\;\,\mathbf{S}_\text{J}$ } } \!\!
                 \end{array}
                \right).
\end{equation}


Physically, the fact that the three components are connected with two ports of the junction each terminated by a qubit makes the 6-by-6 model redundant.  The right going field entering the junction at $z_1$, $F_{+}(z_1)$, is simply the spatial translation of the right-going field that exited qubit a, $F_{+}(z_a)$.  The monochromatic approximation of Eq.~\ref{eq:MonoField} implies that so long as $|z_1 - z_a| \ll \ell_0$, this translation is simply a change in phase, $F_{+}(z_1) = e^{i k_0 (z_1 - z_a)}F_{+}(z_a)$, and there is only one relevant field operator, $b_+(t)$.  Thus the connection from $a$ to the junction input at $z_1$ imposes the constraint
\begin{equation}
    b_{1}^{\ms{in}} = e^{i k_0 (z_1 - z_a)} b_a^{\ms{out}}.
\end{equation}
If $\mathbf{S}_{\ms{J}}$ contains retroreflecting amplitudes at positions $z_1$ and $z_3$, then there is the distinct possibility of developing circulating power in the (hopefully weak) cavities formed in the intervals $[z_a, z_1]$ and $[z_3, z_b]$.  The legitimacy of the monochromatic approximation ultimately depends upon these intermediate cavities being of poor quality with a rapid decay rate, see Section~\ref{ss:regime}.

Fig.~\ref{fig:cross} contains a total of 4 constraints, as both right and left waveguides have bidirectional connections.  For a general network consisting of $N$ inputs and $N$ outputs, if $M \le N$ of the outputs are injectively routed to $M$ distinct inputs then the $M$ constraint equations can be easily written as a single matrix equation relating $M$ elements of $\mathbf{b}_{\ms{out}}(t)$ to $M$ elements of $\mathbf{b}_{\ms{in}}(t)$.

We will find it extremely convenient to utilize matrix projectors that isolate the $M$ connected ``internal'' modes from the remaining $N - M$ free ``external'' modes.  In addition to distinguishing internal from external, we must also maintain the distinction between input and output modes, as they refer to traveling-wave fields in physically distinct regions.  Thus we define the projectors onto the internal/external \emph{inputs} as $\mathbf{I}_{\ms{i}}$ and $\mathbf{X}_{\ms{i}}$ and the projectors onto the internal/external \emph{outputs} as $\mathbf{I}_{\ms{o}}$ and $\mathbf{X}_{\ms{o}}$.  As they are orthogonal projectors they satisfy the relations
\begin{equation}
   \ident = \mathbf{I}_{\ms{i}} + \mathbf{X}_{\ms{i}}, \quad \mathbf{X}_{\ms{i}}^2 = \mathbf{X}_{\ms{i}} , \quad \mathbf{I}_{\ms{i}} \mathbf{X}_{\ms{i}} = \mathbf{X}_{\ms{i}} \mathbf{I}_{\ms{i}} = 0,
\end{equation}
where $\ident$ is the $N \times N$ identity matrix.  We denote the internal/external partitioning of the vector of field operators as superscripts, i.e.\
\begin{equation}
   \mathbf{b}_{\ms{in}} = \mathbf{X}_{\ms{i}}\mathbf{b}_{\ms{in}} + \mathbf{I}_{\ms{i}}\mathbf{b}_{\ms{in}} = \mathbf{b}_{\ms{in}}^{\ms{ext}} + \mathbf{b}_{\ms{in}}^{\ms{int}}.
\end{equation}

Given this notation, the constraints imposed by a given network (that is, which of the outputs are routed to which inputs) can be captured by a single connection matrix, $\mathbf{W}$~\footnote{We have chosen to denote the connection matrix by $\mathbf{W}$ because in mathematical terminology it is a ``weighted adjacency matrix''}, via the relation
\begin{equation}
  \mathbf{b}_{\ms{in}}^{\ms{int}} = \mathbf{W} \mathbf{b}_{\ms{out}}^{\ms{int}}.
\end{equation}
For the crossed-waveguide network depicted in Fig.~\ref{fig:cross}, the connection matrix is explicitly
\begin{equation}
\mathbf{W} = \left(
               \begin{array}{cccccc}
                  0 & 0 & e^{i \varphi_{1a}} & 0 & 0 & 0 \\
                  0 & 0 & 0 & 0 & e^{i \varphi_{3b}} & 0 \\
                  e^{i \varphi_{1a}} & 0 & 0 & 0 & 0 & 0 \\
                  0 & 0 & 0 & 0 & 0 & 0 \\
                  0 & e^{i \varphi_{3b}} & 0 & 0 & 0 & 0 \\
                  0 & 0 & 0 & 0 & 0 & 0 \\
               \end{array}
             \right),
\end{equation}
where the propagation phases are $\varphi_{ij} = k_0 |z_i - z_j|$.  Note that $\mathbf{W}$ is symmetric here because we are considering bidirectional (i.e.\ reciprocal) connections, which may not hold for more general networks.  However, the rank of $\mathbf{W}$ is always equal to the number of internal connections, $M$.  In fact we will repeatedly use the relations,
\begin{equation}
  \mathbf{W}^\dagger \mathbf{W} = \mathbf{I}_{\ms{o}}\quad \text{and}\quad \mathbf{W} \mathbf{W}^\dagger = \mathbf{I}_{\ms{i}}.
\end{equation}
It then follows that $\mathbf{W} = \mathbf{I}_{\ms{i}} \mathbf{W} \mathbf{I}_{\ms{o}}$.

It is worth emphasizing that, at its most general, $\mathbf{S}$ is a unitary map that takes all inputs to all outputs. As a matrix, its columns index input modes while its rows index the outputs. In contrast, $\mathbf{W}$ is a one-for-one mapping between a subset of output modes to an equal number of inputs.  So as a matrix, $\mathbf{W}$ has columns labeled by outputs, rows labeled by inputs, and is null on every unconstrained external mode.  In a sense, $\mathbf{W}$ describes $M$ direct feedback connections, as it returns outputs to inputs.

\section{The effective input-output model for a network with weak loops}
\label{sec:effio}

We have shown above how to define a set of network elements (a set of systems and beam-splitters, each with a number of inputs and outputs), and how to specify the way these elements are connected together to form a network. Recall that if our network contains loops there is no guarantee that it can be described with input-output theory. We now derive the effective input-output description for such a network, and the conditions under which this description can be expected to well-approximate the dynamics of the network.

\subsection{Paths of the network} \label{ss:paths}
A great amount of physical intuition can be gained by noting that every possible path through the network can be created by taking alternating products of $\mathbf{S}$ and $\mathbf{W}$.  Consider the infinite sum of matrices
\begin{equation}
  \mathbf{S} + \mathbf{SWS} + \mathbf{SWSWS} + \dots\ .
\end{equation}
We have already discussed that matrix elements $[\mathbf{S}]_{jk}$ are the scattering amplitudes that directly map $b^{\ms{in}}_k$ to $b^{\ms{out}}_j$.  The elements of the second term of the above series, which we can equivalently write as $\left[\mathbf{SI}_{\ms{i}}\mathbf{WI}_{\ms{o}}\mathbf{S}\right]_{jk}$, is the sum over all paths that map input $k$ to output $j$, while traversing the network exactly once.  If $j$ is an internal output then the path will continue and traverse the network a second time.   In general $\left[( \mathbf{SW})^n \mathbf{S}\right]_{jk}$ is the sum over all $k \mapsto j$ paths that traverse the network precisely $n$ times.  Assuming this series converges (that is, longer paths have successively smaller amplitudes) we then concluded that the elements of the matrix
\begin{equation}
 \Biggl[ \sum_{n=0}^{\infty} (\mathbf{S}\mathbf{W})^n \Biggr] \mathbf{S} = \frac{1}{\ident - \mathbf{SW}}\mathbf{S}
\end{equation}
are the coherent sum over all scattering paths through the network.  The subset of these paths that take external inputs to external outputs are thus the nonzero elements of
\begin{equation}
  \mathbf{S}_{\ms{eff}} \equiv \mathbf{X}_{\ms{o}} \frac{1}{\ident - \mathbf{SW}} \mathbf{S} \mathbf{X}_{\ms{i}},
\end{equation}
which is the effective scattering matrix of the reduced input/output model.

The vector of effective sources $\mathbf{L}_{\ms{eff}}$ is interpretable along similar lines.  Consider the sum
\begin{equation}
  \mathbf{L} + \mathbf{SWL} + \mathbf{SWSWL} + \dots\
\end{equation}
This is the coherent sum over the raw system emissions $\mathbf{L}$, the emissions having progressed through one round trip of the network, $\mathbf{SWL}$, the emissions after two round trips, and so on.  Summing this series and projecting it onto the external outputs gives,
\begin{equation}
  \mathbf{L}_{\ms{eff}} \equiv \mathbf{X}_{\ms{o}} \frac{1}{\ident - \mathbf{SW}} \mathbf{L}.
\end{equation}
Thus we expect that
\begin{equation}
  \mathbf{b}_{\ms{out}}^{\ms{ext}} = \mathbf{S}_{\ms{eff}} \mathbf{b}_{\ms{in}}^{\ms{ext}} + \mathbf{L}_{\ms{eff}}.
\end{equation}

\subsection{The effective Hamiltonian} \label{ss:Heff}

We now turn to identifying how the connection constraints effect the evolution of the local systems embedded in the network.  We will begin by considering the Heisenberg-Langevin equation for the evolution of any system operator $A$, interacting with $N$ independent fields;
\begin{equation}
\label{eq:HLeqs}
\begin{split}
    \dot{A} &= \frac{i}{\hbar}\left[H_{\ms{sys}}, A \right]  - \frac{1}{2} \left(  [A, \mathbf{L}^\dag] \mathbf{L} - \mathbf{L}^\dag [A,\mathbf{L}] \right) \\
  &\quad - [A,\mathbf{L}^\dag ] \mathbf{S} \mathbf{b}_{\ms{in}} + \mathbf{b}_{\ms{in}}^\dag \mathbf{S}^\dag [A, \mathbf{L} ].
\end{split}
\end{equation}
Here we have included an unspecified system Hamiltonian $H_{\ms{sys}}$, to allow for the presence of external controls.  We assume that $H_{\ms{sys}}$, when transformed to the interaction picture, causes slow evolution of the system(s), ensuring that the system-field coupling remains quasi-resonant.  The compact but possibly ambiguous vector notation introduced above should be interpreted as implicit sums:
\begin{equation}
\begin{split}
     [A, \mathbf{L}^\dag] \mathbf{L} & \equiv \sum_i [A, L_i^\dag] L_i , \\
     [A,\mathbf{L}^\dag] \mathbf{S} \mathbf{b}_{\ms{in}} & \equiv \sum_{ij} [A, L^\dag_i ] s_{ij} b^{\ms{in}}_j .
\end{split}
\end{equation}
The presence of the scattering matrix $\mathbf{S}$ in the equations of motion for the system comes from the fact that $L_i$ is coupled to the $i^{\ms{th}}$ output and $\sum_j s_{ij} b^{\ms{in}}_j$ is the free output field for that mode. (Note that this equation fails to hold when $[A, \mathbf{S}] \ne 0$.)

In the appendix we show that by imposing the connection constraints, $\mathbf{b}^{\ms{int}}_{\ms{in}} = \mathbf{Wb}^{\ms{int}}_{\ms{out}}$, the equation of motion for $A$, Eq.(\ref{eq:HLeqs}), is transformed not only by the replacement $(\mathbf{S}, \mathbf{L}) \mapsto (\mathbf{S}_{\ms{eff}}, \mathbf{L}_{\ms{eff}})$, but also by the replacement of the total Hamiltonian, $H_{\ms{sys}}$, with the effective Hamiltonian
\begin{equation} \label{eq:Heff}
  H_{\ms{eff}} \equiv H_{\ms{sys}} + \frac{\hbar}{2i} \mathbf{L}^\dag \left(\frac{\ident}{\ident - \mathbf{SW}} - \frac{\ident}{\ident - (\mathbf{SW})^\dag}\right) \mathbf{L}.
\end{equation}
The proof of this result is essentially an exercise in matrix algebra. Physically the second term in $H_{\ms{eff}}$ is equal to the ``imaginary part'' of the coherent sum over all paths where a quanta is emitted via $\mathbf{L}_j$ and subsequently absorbed by $\mathbf{L}^\dag_{i}$.  Such terms account for the spin exchange rates for atoms coupled to a 1D wave guide, as well as the Lamb shift in either free space~\cite{asenjo-garcia_atom-light_2017} or cavity like~\cite{yao_ultrahigh_2009} conditions.

\subsection{The effective input-output description} \label{ss:IOmodel}

We can now write down the complete effective input-output description for a network that contains ``weak loops'' for the fields. For ease of reference we now collect all the equations that define this effective description. Given that $\mathbf{W}$ is the matrix that defines the connections between the systems (the network topology), and that $\mathbf{b}_{\ms{in}}^{\ms{ext}}$ and $\mathbf{b}_{\ms{out}}^{\ms{ext}}$ are, respectively, the external inputs and outputs to this network, we have
\begin{equation}
  \mathbf{b}_{\ms{in}}^{\ms{ext}} = ( \ident - \mathbf{W}\mathbf{W}^\dagger)  \mathbf{b}_{\ms{in}} ,
\end{equation}
where $\ident$ is the $N$-dimensional identity matrix, and
\begin{eqnarray}
  \mathbf{b}_{\ms{out}}^{\ms{ext}} & = & \mathbf{S}_{\ms{eff}} \mathbf{b}_{\ms{in}}^{\ms{ext}} + \mathbf{L}_{\ms{eff}}
\end{eqnarray}
  with
\begin{eqnarray}
  \mathbf{L}_{\ms{eff}} & \equiv & \mathbf{X}_{\ms{o}} \frac{1}{\ident - \mathbf{SW}} \mathbf{L}, \\
   \mathbf{S}_{\ms{eff}} & \equiv & \mathbf{X}_{\ms{o}} \frac{1}{\ident - \mathbf{SW}} \mathbf{S} \mathbf{X}_{\ms{i}} .
\end{eqnarray}
The effective equation of motion for any system operator $A$ is
\begin{equation}\label{eq:EOMeff}
\begin{split}
  \dot{A} &=  \frac{i}{\hbar}\left[H_{\ms{eff}}, A \right]  - \frac{1}{2} \left(  [A, \mathbf{L}^\dag_{\ms{eff}}] \mathbf{L}_{\ms{eff}} - \mathbf{L}^\dag_{\ms{eff}} [A,\mathbf{L}_{\ms{eff}}] \right)\\
  & \quad - [A,\mathbf{L}^\dag_{\ms{eff}} ] \mathbf{S}_{\ms{eff}} \mathbf{b}^{\ms{ext}}_{\ms{in}} + \left( \mathbf{S}_{\ms{eff}}\mathbf{b}^{\ms{ext}}_{\ms{in}}\right)^\dag [A, \mathbf{L}_{\ms{eff}} ] ,
\end{split}
\end{equation}
in which the effective Hamiltonian is given above in Eq.(\ref{eq:Heff}).

\subsection{The non-hermitian dissipative ``Hamiltonian''} \label{ss:Hdiss}
While the operators $H_{\ms{eff}}$ and $\mathbf{L}_{\ms{eff}}$ are all that are needed to specify the master equation in Lindblad form (for vacuum input fields), some physical insight can be gained by considering the non-hermitian dissipative ``Hamiltonian'', $\widetilde{H}_{\ms{loss}} \equiv  H_{\ms{eff}} - i \frac{1}{2}\mathbf{L}_{\ms{eff}}^\dag \mathbf{L}_{\ms{eff}}$.  An application of the tricky relation in Eq.(\ref{eq:app:trick2}) shows that,
\begin{equation}
\begin{split}
    \widetilde{H}_{\ms{loss}} &= H_{\ms{sys}} - \tfrac{i}{2} \mathbf{L}^\dag  \mathbf{L} - i \mathbf{L}^\dag \tfrac{\mathbf{SW}}{\ident - \mathbf{SW}} \mathbf{L}.
\end{split}
\end{equation}
Each term of $\widetilde{H}_{\ms{loss}}$ has a clear and distinct meaning.  The first and second show the ``unconstrained'' open system dynamics, consisting of the externally applied $H_{\ms{sys}}$ and the dissipation induced by every output port.  The third term shows the cumulative effect of the network.  The coherent effects of $H_{\ms{eff}}$ and the dissipation induced by $\mathbf{L}_{\ms{eff}}^\dag \mathbf{L}_{\ms{eff}}$ combine as real and imaginary parts so that $\widetilde{H}_{\ms{loss}}$ only involves the sum $\mathbf{SW} + \mathbf{SWSW} + \dots $ and not its ``reversed'' adjoint.  If $\mathbf{W}$ is asymmetric and only connects system $a$ to system $b$ but not vice-versa, then this unidirectional flow is preserved in $\widetilde{H}_{\ms{loss}}$.

\subsection{Regime of applicability: satisfying weak-coupling and multipass constraints} \label{ss:regime}

\begin{figure}[t]
    \begin{center}
    \includegraphics[width=0.95\hsize]{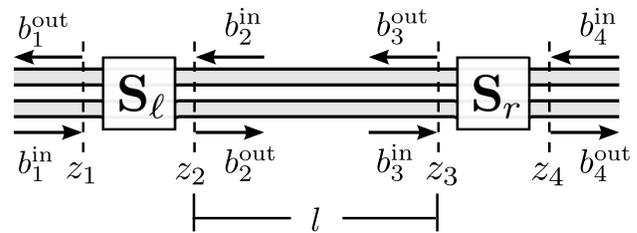}
        \caption{\label{fig:FabryPerot} An empty cavity.  A section of waveguide of length $z_3 - z_2 = l$ is capped by two partially reflecting boundary conditions: $\mathbf{S}_\ell$ and $\mathbf{S}_r$.}
	\end{center}
\end{figure}

Here we identify the conditions under which the effective input-output description, derived above, is a good approximation to the real network in which the fields pass through the loops multiple times. This involves identifying when the network geometry allows repeated interactions, while ensuring that the criteria for the monochromatic traveling-wave approximation remains valid. Consider the case depicted in Fig.~\ref{fig:FabryPerot} where a simple empty cavity is formed when a section of waveguide of length $l$ is capped by two partially reflecting boundary conditions.  The scattering elements at these boundaries are defined by the frequency independent unitary matrices $\mathbf{S}_\ell$ and $\mathbf{S}_r$.  In the monochromatic approximation, the in-to-out scattering boundary condition, and the constraint of free propagation of internal modes are imposed by the block diagonal matrices,
\begin{equation}
  \mathbf{S}  =   \left( \scalebox{1.25}{ $
                  \begin{array}{cc}
                     \mathbf{S}_\ell & \mathbf{0} \\
                     \mathbf{0} & \mathbf{S}_r \\
                  \end{array}
                $ } \right)
\end{equation}
and
\begin{equation}
  \mathbf{W}  = \left(\scalebox{0.8}{ $
                  \begin{array}{cccc}
                    0 & 0 & 0 & 0 \\
                    0 & 0 & e^{i k_0 l} & 0 \\
                    0 & e^{i k_0 l} & 0 & 0 \\
                    0 & 0& 0 & 0 \\
                  \end{array} $ }
                \right)
\end{equation}

Outside of this approximation, we still have the two parameter field operators $b_\pm(z_i, t)$, which can still be arranged into input/output vectors $\mathbf{b}_{\ms{in}}(t)$ and $\mathbf{b}_{\ms{out}}(t)$.

The free-propagation of the internal fields still give the constraints, 
\begin{equation}
\begin{split}
 F_{+}(z_3, t)  &= F_{+}(z_2, t - l/v_p)\\
 F_{-}(z_2,t) &= F_{-}(z_3, t - l/v_p).
\end{split}
\end{equation}
This implies that
\begin{equation}
 \mathbf{b}^{\ms{int}}_{\ms{in}}(t) = \mathbf{Wb}^{\ms{int}}_{\ms{out}}(t - l/v_p).
\end{equation}
Combining this with the general condition
\begin{equation}
 \mathbf{b}_{\ms{out}}(t) = \mathbf{Sb}_{\ms{in}}(t) + \mathbf{L}(t)
\end{equation}
shows that,
\begin{equation}\label{eq:delayedout}
\begin{split}
  \mathbf{b}_{\ms{out}}(t) &= \sum_{n = 0}^{\infty} (\mathbf{SW})^n \,\mathbf{SX}_{\ms{i}}\mathbf{b}_{\ms{in}}(t - n l/v_p) \\
   &\quad + \sum_{n = 0}^{\infty} (\mathbf{SW})^n \, \mathbf{L}(t - n l/v_p).
\end{split}
\end{equation}
Note that this is simply the time-delayed version of Eq.(\ref{eq:app:Boutsubs}).

Previously we have argued that so long as the spatial separation remains small when compared to the characteristic length $\ell_0 = v_p/\kappa_0$ then the field operators are still well approximated by a single field operator that delta commutes in time.  But here we have an infinite number of distances so that there always exists a number reflections $n_{\ms{cut}}$ such that $n_{\ms{cut}} l \sim \ell_0$.   Nevertheless, the contribution of these significantly delayed paths are all attenuated by a factor of $(\mathbf{SW})^{n_{\ms{cut}}}$.  If the largest singular value of this matrix is small compared to 1, then we can be confident in the approximation.

In a general network the time delay for the different internal connections may be different. Rather than being merely $n$ multiples of a single traversal time, the cumulative delay will depend upon the specific path taken.  For a given path consisting of $n$ internal connections followed by a final exit at an external port, the total delay $\tau_n$ is simply the sum over the individual delays, so that
\begin{equation}
  \tau_n = \sum_{i} |z_{i+1} - z_{i}|/v_p
\end{equation}
and the weighting factor $w_n$ is the product of matrix elements
\begin{equation}
  w_n = [\mathbf{SW}]_{e,i_n}[\mathbf{SW}]_{i_n, i_{n-1}}\cdots [\mathbf{SW}]_{i_2, i_1} .
\end{equation}
Ultimately, we require that any path that has a significant delay must also have an insignificant weight. We can therefore state a simple sufficient condition for the validity of the effective description as follows.  If $\tau_{\ms{min}}$ is the minimum relevant system dynamical timescale (e.g.\ $1/\kappa_0$) then
\begin{equation}
    w_n \ll 1 \quad \text{for all}\quad \tau_n \gtrsim \tau_{\ms{min}}.
\end{equation}
This condition captures the requirement that any effective cavities that are formed by loops in the network decay sufficiently fast, as discussed in Section~\ref{mimo}. 

For example, consider the two qubit system of Fig~\ref{fig:reductions}(c), where two qubits are connected by a transmission line, via isolating circulators.  Taking this as a model for a long distance communication channel, let the interconnecting distance, $L = |z_4 - z_2|$, be generally larger than the distances between the bare qubits and their circulators, $\Delta z_{a} = |z_1 - z_a|$ and $\Delta z_{b} = |z_5 - z_b|$.  The paths with the longest delay times are clearly those that traverse the interconnecting line.  The shortest path traveling from qubit $a$ to qubit $b$ follows the sequence of outputs with locations $z_a \rightarrow z_2 \rightarrow z_5 \rightarrow z_b$.  This path has a delay time $\tau_1 = (L + \Delta z_a + \Delta z_b)/v_p$ and carries a weight with magnitude $|w_1| = |t_{54}t_{21}|$.  The paths following outputs  $z_a \rightarrow z_2 \rightarrow z_4 \rightarrow z_1 \rightarrow z_a$, and $z_a \rightarrow z_2 \rightarrow z_4 \rightarrow z_2 \rightarrow z_5 \rightarrow z_b$ have delays of $\tau_2 = 2(L + \Delta z_a)/v_p$ and $\tau_3 = (3 L + \Delta z_a + \Delta z_b)/v_p$.  The corresponding weights have magnitudes $|w_2| = |c_{12}r_{44}t_{21}|$ and $|w_3| = |t_{54}r_{22}r_{44}t_{21}|$.  An off-the-shelf circulator operating in the 4-8 GHz range (e.g. \emph{Low Noise Factory - LNF-ISC4\_8A}), specifies the parameters $|t_{ij}| \sim 0.98$, and $|r_{ij}|, |c_{ij}| \lesssim 0.08$.  Thus for the delays $\tau_1, \tau_2, \tau_3$ the associated weights have magnitudes $|w_1| \sim 0.96$, and $|w_2|, |w_3| \sim 6 \times 10^{-3}$.

Were we to follow the manufacturer's lead and consider the weights $|w_2|$ and $|w_3|$ negligible, we then recover the standard implementation of cascading two systems in a unidirectional way.  In this case, the time-of-flight delay, $\tau_1$ can be easily absorbed by evaluating qubit $b$ at the retarded time $t_r = t - \tau_1$~\cite{gardiner_driving_1993}.

\section{An example: the qubit-to-qubit communication channel}
\label{sec:example}

One of the many uses of a quantum communication channel is the transfer of entanglement.  Specifically, Alice wishes to send to Bob one half of an entangled state.  Previous results~\cite{wenner_catching_2014, Srinivasan14, yin_catch_2013, Bader13, korotkov_flying_2011, jahne_high-fidelity_2007, cirac_quantum_1997, Axline18, Kurpiers18} show how one could transfer a quantum state between localized systems by dynamically controlling the rates each system couples to a one dimensional traveling-wave field.  In a lossless setting, near perfect transfer fidelity can be achieved if the sender and receiver use appropriately matched waveforms.  For example, if the receiver, Bob, simply gates his coupling in an on/off way, the sender Alice should modulate her coupling rate such that the outgoing wave packet forms a rising exponential, effectively the time reversal of an excited system decaying at a constant rate~\cite{korotkov_flying_2011}.  Here we investigate if and how such a procedure could be implemented across an imperfect network.

For simplicity we assume that all input fields are in the vacuum state, which is not an unreasonable idealization, for example, for superconducting qubits with $\omega_0 \sim 6 \text{ GHz}$ and an ambient temperature no higher than $100 \text{ mK}$.  Fig.~\ref{fig:reductions}(c) and Fig.~\ref{fig:cross} depict two different, yet similar settings where two qubits are interconnected via guided fields with two external inputs and two external outputs.   The design in Fig.~\ref{fig:reductions}(c) models an asymmetric communication channel where two qubits are linked via a transmission line and are isolated by circulators that may be imperfect. The design in Fig.~\ref{fig:cross} models two qubits connected to transmission lines that cross at a beam-splitter-like intersection. Nevertheless, depending upon how the scatter $\mathbf{S}_{\ms{J}}$ in the latter couples its four ports, both models may generate the same dynamics for the qubits. The difference between the two really lies in the quality of the zero-delay approximation, where a significant delay in the interconnecting line may violate the necessary assumptions.

In either of the above models we assume that local, possibly time-dependent rotations can be applied to each qubit, so that
\begin{equation}
  H_{\ms{sys}} = \frac{\hbar}{2} \, \left[ \mathbf{h}_a(t)\cdot\boldsymbol{\sigma}_{a} + \mathbf{h}_b(t)\cdot\boldsymbol{\sigma}_{b}\,\right],
\end{equation}
where $\mathbf{h}_a$ is the control vector and $\boldsymbol{\sigma}_{a}$ is the vector of Pauli matrices for qubit $a$; and likewise for qubit $b$. We also assume that each qubit couples only to its transmission line and not to any additional output. We imagine that the decay rates of both qubits, $\kappa_i$, and their coupling phase angles $\phi_i$, can be independently controlled, so that
\begin{equation}
  \mathbf{L} = \left\{\begin{array}{cc}
                       \sqrt{\kappa_i(t)}\,  e^{i \phi_i(t) }\, \sigma^{-}_i & \text{ for } i \in \{a,b\} \\
                       0 & \text{ otherwise}
                     \end{array}\right. .
\end{equation}

To apply our effective network theory to the network connecting the two qubits it is useful to recall first that there is a single matrix that plays the key role in determining the effective input-output description, namely
\begin{equation}
  \frac{1}{\ident - (\mathbf{SW})} = \ident + \frac{\mathbf{SW}}{\ident - (\mathbf{SW})}.
\end{equation}
It is this matrix that tells how the outputs of the systems are mapped (routed, if you like) to the external outputs of the network. Examining the expression for the effective Lindblad operators,
\begin{equation}
  \mathbf{L}_{\ms{eff}} = \mathbf{X}_{\ms{o}} \frac{1}{\ident - \mathbf{SW}} \mathbf{L} = \mathbf{X}_{\ms{o}} \left[ \mathbf{L} +  \frac{\mathbf{SW}}{\ident - \mathbf{SW}} \mathbf{L} \right],
\end{equation}
we see that it is the matrix
\begin{equation} \label{eq:defT}
  \mathbf{T} \equiv \frac{\mathbf{SW}}{\ident - (\mathbf{SW})}
\end{equation}
that gives the contribution of the network, since it is this matrix that vanishes when $W=0$, in which case $\mathbf{L}_{\ms{eff}}$ reduces to $\mathbf{X}_{\ms{o}} \mathbf{L}$.

With some abuse of notation, we will denote the matrix element of $\mathbf{T}$ that maps the output from qubit $a$ or $b$ to the $j^{\msi{th}}$ output by
\begin{equation}
  t_{j k} \equiv \left[ \mathbf{T} \right]_{jk} = \left[\tfrac{\mathbf{SW}}{\ident - \mathbf{SW} } \right]_{jk}, \;\; \mbox{with}\;\;\; k = a,b.
\end{equation}
Using these matrix elements we can now write explicit expressions for the effective Lindblad operators (the elements of $\mathbf{L}_{\ms{eff}}$). The Lindblad jump operator associated with a photon leaving external port $j$ is
\begin{equation}
  \begin{split}
    L_{\ms{eff}\, j} =& t_{ja} L_a + t_{jb} L_b,
  \end{split}
\end{equation}
and the effective Hamiltonian is ( $\hbar = 1$ )
\begin{equation}
\begin{split}
    H_{\ms{eff}} &= H_{\ms{sys}}
     + L_a^\dag\, \operatorname{Im}( t_{aa} ) L_a  + L_b^\dag \operatorname{Im}( t_{bb} ) L_b \\
    &\quad + \tfrac{1}{2i} L_a^\dag\, ( t_{ab} - t_{ba}^*)\, L_b  + \tfrac{1}{2i} L_b^\dag\, ( t_{ba} - t_{ab}^*)\, L_a.
\end{split}
\end{equation}

Given vacuum external inputs to the network, the master equation for the two-qubit density matrix $\rho$, when written in Lindblad form is
\begin{equation}
\begin{split}
    \dot{\rho} & = -i[ H_{\ms{eff}}, \rho] +\mathcal{D}[\mathbf{L}_{\ms{eff}}]\rho
\end{split}
\end{equation}
in which the super-operator $\mathcal{D}[\mathbf{L}_{\ms{eff}}]\rho$ is defined by
\begin{equation}
   \mathcal{D}[\mathbf{L}_{\ms{eff}}]\rho \equiv \mathbf{L}_{\ms{eff}}\, \rho\, \mathbf{L}_{\ms{eff}}^\dag - \tfrac{1}{2} \mathbf{L}_{\ms{eff}}^\dag \mathbf{L}_{\ms{eff}}\, \rho - \tfrac{1}{2} \rho \, \mathbf{L}_{\ms{eff}}^\dag \mathbf{L}_{\ms{eff}}.
\end{equation}

The derivation of $H_{\ms{eff}}$ in Appendix~\ref{app:algebra} shows that $\mathbf{L}_{\ms{eff}}^\dag \mathbf{L}_{\ms{eff}}$ can be simplified, via Eq.(\ref{eq:app:trick}), to read,
\begin{equation}
\begin{split}
  \mathbf{L}_{\ms{eff}}^\dag \mathbf{L}_{\ms{eff}} &= \mathbf{L}^\dag \left( \ident + \frac{\mathbf{SW}}{\ident - \mathbf{SW}} + \frac{(\mathbf{SW})^\dag}{\ident - (\mathbf{SW})^\dag} \right) \mathbf{L} \\
  &= \sum_{i,j \in \{a,b\}} L^\dag_i \left(\delta_{ij} + t_{ij} + t_{ji}^* \right) L_{j}.
\end{split}
\end{equation}
Additionally, the fact that the matrix elements $t_{ij}$ are merely complex numbers and thus commute with all system operators means that
\begin{equation}
\begin{split}
  \mathbf{L}_{\ms{eff}}\, \rho\, \mathbf{L}_{\ms{eff}}^\dag &= \sum_{k \in \substack{ \ms{ext}\\ \ms{outs}} } \sum_{j,i \in \{a,b\}}  t_{kj}  L_j\,  \rho \, t_{k i}^* L_i^\dag \\
   &=  \sum_{j,i \in \{a,b\}} \Big[ \sum_{k \in \substack{ \ms{ext}\\ \ms{outs}} } t_{k i}^* t_{kj} \Big] L_j\, \rho \, L_i^\dag \\
   &=  \sum_{j,i \in \{a,b\}} \left[ \tfrac{1}{\ident - (\mathbf{SW})^\dag}\mathbf{X}_{\ms{o}} \tfrac{1}{\ident - \mathbf{SW}} \right]_{ij} L_j\, \rho \, L_i^\dag \\
   &=  \sum_{j,i \in \{a,b\}} ( \delta_{ij} + t_{ij} + t_{ji}^* ) L_j\, \rho \, L_i^\dag.
\end{split}
\end{equation}
Combining these two results we find that the two-qubit master equation can also be written as
\begin{eqnarray}
     \dot{\rho} & = & -\frac{i}{2}\Big[  \mathbf{h}_a \cdot\boldsymbol{\sigma}_{a} + \mathbf{h}_b \cdot\boldsymbol{\sigma}_{b}  - i \!\!\!\!\! \sum_{i,j \in \{a,b\}} \!\!\!\! L_i^\dag \left(t_{ij} - t_{ji}^* \right) L_j,\, \rho \Big] \nonumber \\
     & & + \! \sum_{i,j \in \{a,b\}}\!\!\!\! \left(\delta_{ij} + t_{ij} + t_{ji}^* \right) \left( L_{j}\, \rho\, L_i^\dag - \tfrac{1}{2} \left\{ L^\dag_i L_{j},  \rho \right\} \right) \nonumber \\
\end{eqnarray}
where $\{A,B\} \equiv AB + BA$ is the anti-commutator.

This master equation can exhibit super/sub-radiant states, where interference between the various decay channels results in distinct decay rates for different superpositions of the two atom states.  There may even be a specific superposition of the states $\left\vert \uparrow \downarrow\right\rangle$ and $\left\vert \downarrow \uparrow\right\rangle$ which is unable to radiate and remains a dark state.  If such a state exists and its amplitudes are suitably controllable then it could be used to deterministically transfer a single excitation across the network.  Next we show that such a state exists if and only if
\begin{equation} \label{eq:perfectConstraint}
  \| t_{ab}^* + t_{ba} \|^2 = (1 + t_{aa} + t_{aa}^*)(1 + t_{bb} + t_{bb}^*).
\end{equation}

\subsection{Two-qubit superradience}
\label{ss:suprad}

We identify the emission properties of the two-qubit system by first analyzing how the populations of different states are transferred via the so called ``feeding terms'' of the master equation.  Specifically, consider the expression
\begin{equation}
  \sum_{i,j \in \{a,b\}} \left(\delta_{ij} + t_{ij} + t_{ji}^* \right) L_{j}\, \rho\, L_i^\dag.
\end{equation}
The single-qubit terms (i.e.\ those for which $i = j$) show that the network has the effect of increasing the action of these terms by the factor
\begin{equation}
  \eta_i \equiv  1 + t_{ii} + t_{ii}^*\quad \text{ for } i = a,b,
\end{equation}
which is the 1D equivalent of the Purcell factor.

A straightforward calculation shows that when written as outer products of the two-qubit basis states $\{ \left\vert \uparrow \downarrow \right\rangle, \left\vert\downarrow \uparrow \right\rangle,  \left\vert \downarrow \downarrow \right\rangle, \left\vert \uparrow \uparrow \right\rangle\}$,
\begin{multline}\label{eq:feeding}
  \sum_{i,j \in \{a,b\}} \left(\delta_{ij} + t_{ij} + t_{ji}^* \right) L_{j}\, \rho\, L_i^\dag = \\
   \operatorname{Tr} \left(\, \rho \left\vert \uparrow \uparrow \right\rangle\! \left\langle \uparrow \uparrow  \right\vert\, \right) R
    + \operatorname{Tr} \left( R \rho \right) \left\vert \downarrow \downarrow \right\rangle\! \left\langle \downarrow \downarrow  \right\vert
\end{multline}
where
\begin{multline}
R \equiv \kappa_a\, \eta_a \left\vert \uparrow \downarrow \right\rangle\! \left\langle \uparrow \downarrow \right\vert + \kappa_b\, \eta_b \left\vert \downarrow \uparrow  \right\rangle\! \left\langle \downarrow \uparrow \right\vert\\
+ \sqrt{\kappa_a \kappa_b} (t_{ab}^* + t_{ba})\,e^{i \phi_a - i \phi_b} \left\vert \downarrow \uparrow  \right\rangle\! \left\langle \uparrow \downarrow \right\vert\\
+ \sqrt{\kappa_a \kappa_b} (t_{ab} + t_{ba}^*)\,e^{- i \phi_a + i \phi_b} \left\vert \uparrow \downarrow \right\rangle\! \left\langle \downarrow \uparrow \right\vert.
\end{multline}
In terms of $R$ the total master equation is
\begin{equation}\label{eq:masterEq}
\begin{split}
    \dot{\rho} &= -i[ H_{\ms{eff}}, \rho] + \operatorname{Tr} \left(\, \rho \left\vert \uparrow \uparrow \right\rangle\! \left\langle \uparrow \uparrow  \right\vert\, \right) R + \operatorname{Tr} \left( R \rho \right) \left\vert \downarrow \downarrow \right\rangle\! \left\langle \downarrow \downarrow  \right\vert \\
    &\quad - \tfrac{1}{2}\operatorname{Tr} \left( R \right)\, \big( \left\vert \uparrow \uparrow \right\rangle\! \left\langle \uparrow \uparrow \right\vert \rho + \rho \left\vert \uparrow \uparrow \right\rangle\! \left\langle \uparrow \uparrow  \right\vert \big) - \tfrac{1}{2} \left( R \rho + \rho R \right).
\end{split}
\end{equation}
By writing it in this way, we see that the properties of $R$ not only characterize the decay of the single-excitation subspace (being the subspace in which one but not both of the qubits is in its excited state), but also the decay rate of the doubly-excited state $\left\vert \uparrow \uparrow \right\rangle$ and the accumulation rate of the zero excitation state $\left\vert \downarrow \downarrow \right\rangle$.

In the single-excitation subspace, $\operatorname{span}\, \{ \left\vert \uparrow \downarrow \right\rangle, \left\vert \uparrow \downarrow \right\rangle \}$, $R$ forms a $2 \times 2$ matrix, which is easily diagonalized.  Looking forward to a specific application, we find it useful to introduce a Bloch-sphere representation in this 2-level subspace. By choosing the representation $|0\rangle \equiv \left\vert \uparrow \downarrow \right\rangle$ and $|1\rangle \equiv \left\vert \downarrow \uparrow\right\rangle$, the usual 4 component Pauli matrices $(I, \sigma_x, \sigma_y, \sigma_z)$ span the space of $2 \times 2$ complex matrices.  Thus in this representation,
\begin{equation}
  R = \tfrac{1}{2}\left(\mathsf{R}_0 I + \mathsf{R}_x \sigma_x + \mathsf{R}_y \sigma_y + \mathsf{R}_z \sigma_z \right)
\end{equation}
where
\begin{equation}\label{eq:R}
  \begin{split}
    \mathsf{R}_0 &= \kappa_a\, \eta_a + \kappa_b\, \eta_b \\
    \mathsf{R}_x &= 2 \sqrt{\kappa_a \kappa_b}\, \beta_+\, \cos( \phi_a - \phi_b + \delta_{+} ) \\
    \mathsf{R}_y &= 2 \sqrt{\kappa_a \kappa_b}\, \beta_+\, \sin( \phi_a - \phi_b + \delta_{+} ) \\
    \mathsf{R}_z &= \kappa_a\, \eta_a - \kappa_b\, \eta_b ,
  \end{split}
\end{equation}
and with a bit of foresight it is useful to define the cross-coupling coefficients
\begin{equation}
    \beta_{\pm} \equiv \left\| t_{ab}^* \pm t_{ba}  \right\|
\end{equation}
and associated phase angles
\begin{equation}
    \delta_{\pm} \equiv \arg ( t_{ab}^* \pm t_{ba}  ).
\end{equation}
In this notation, the eigenvalues of $R$ define the collective decay rates
\begin{equation}
\begin{split}
    \Gamma_{b/d} &= \frac{1}{2} \left( R_0 \pm \sqrt{ R_x^2 + R_y^2 + R_z^2}\, \right)\\
      &= \frac{1}{2} \left( R_0 \pm \sqrt{ 4 \kappa_a \kappa_b \beta_+^2 + (\kappa_a \eta_a - \kappa_b \eta_b )^2 }\, \right) \! .
\end{split}
\end{equation}
The corresponding eigenstates are the superradiant bright state $|B\rangle$ and the subradiant dark state $|D\rangle$:
\begin{equation}
\begin{split}
    |B\rangle &= \cos(\vartheta/2)|0\rangle + \sin(\vartheta/2) e^{i \varphi } |1\rangle \quad \text{ and }\\
    |D\rangle &= \sin(\vartheta/2)|0\rangle - \cos(\vartheta/2) e^{i \varphi } |1\rangle,
\end{split}
\end{equation}
where $\varphi = \phi_a - \phi_b + \delta_+$ and the mixing angle $\vartheta$ satisfies
\begin{equation}
  \tan\, \vartheta = \frac{2 \sqrt{\kappa_a \kappa_b} \beta_+}{ \kappa_a\, \eta_a - \kappa_b\, \eta_b}.
\end{equation}
Other than the trivial solution $\kappa_a = \kappa_b = 0$, the only way for $|D\rangle$ to be a truly dark state with $\Gamma_d = 0$ is when $\beta_+^2 = \eta_a \eta_b$, i.e.\ when Eq.(\ref{eq:perfectConstraint}) is satisfied.  One possible configuration that meets this criteria is the perfect unidirectional communication channel, e.g.\ when $t_{aa} = t_{bb} = t_{ab} = 0$ and $|t_{ba}| = 1$.

The key insight is that sweeping from a parameter regime in which $0 < \kappa_a \ll \kappa_b $ to that in which $0 < \kappa_b \ll \kappa_a $ results in sweeping $\vartheta$ from $\pi$ to $0$. This in turn sweeps the state $|D\rangle$ from $|0\rangle$ to $|1\rangle$.  Thus if the remaining coherent terms of the overall master equation can be engineered so that the total system evolution also follows $|D\rangle$ then a single excitation can be transferred from the first qubit to the second with a minimum amount of radiative loss.

\subsection{The single-excitation Bloch vector}
\label{ss:bloch}

Here we show that there exists a control scheme such that the joint system will evolve from the state $\left\vert \uparrow \downarrow \right\rangle$ to the state $\left\vert \downarrow \uparrow \right\rangle$ while simultaneously maximizing the overlap of the evolving state with the subradiant state $|D\rangle$. Note that both the state of the system and the subradiant state change with time as the control parameters change with time. The initial state is thus $\rho(0) = \left\vert \uparrow \downarrow \right\rangle \! \left\langle \uparrow \downarrow \right\vert$.

Here we will exclusively consider local qubit rotation vectors $\mathbf{h}_a(t)$ and $\mathbf{h}_b(t)$ that only induce $\sigma_z$ rotations.  The idea being that if the total number of excitations in the system is a conserved quantity, then the states $\left\vert \uparrow \downarrow \right\rangle$ and $\left\vert \downarrow \uparrow \right\rangle$ will form a closed subspace.  In order for a given observable to be a constant of motion, and thereby conserved, it must commute with the total Hamiltonian.  But as $\sigma^{(a)}_x$ changes the number of excitations in qubit $a$ irrespective of qubit $b$, a control Hamiltonian that contains single qubit $x$ or $y$ rotations cannot preserve the total number of excitations.

Given this constraint, we write this excitation-number conserving $H_{\ms{eff}}$ in the two-qubit basis, which is,
\begin{equation}
\begin{split}
    H_{\ms{eff}} &= \left(\kappa_a \operatorname{Im} (t_{aa}) + \kappa_b \operatorname{Im}(t_{bb}) + \tfrac{1}{2} {h_a}_z  + \tfrac{1}{2} {h_b}_z \right) \left\vert \uparrow \uparrow \right\rangle\! \left\langle \uparrow \uparrow  \right\vert\\
    &\quad - \left(\tfrac{1}{2} {h_a}_z  + \tfrac{1}{2} {h_b}_z \right) \left\vert \downarrow \downarrow \right\rangle\! \left\langle \downarrow \downarrow  \right\vert \\
    &\quad+ \left(\kappa_a \operatorname{Im} (t_{aa}) + \tfrac{1}{2} {h_a}_z  -  \tfrac{1}{2} {h_b}_z \right) \left\vert \uparrow \downarrow \right\rangle\! \left\langle \uparrow \downarrow  \right\vert \\
    &\quad+ \left(\kappa_b \operatorname{Im}(t_{bb}) - \tfrac{1}{2} {h_a}_z  +  \tfrac{1}{2} {h_b}_z \right) \left\vert \downarrow \uparrow \right\rangle\! \left\langle \downarrow \uparrow  \right\vert \\
    &\quad+\tfrac{1}{2i} \sqrt{\kappa_a \kappa_b} \|t_{ab}^* - t_{ba}\| e^{-i (\phi_a - \phi_b + \delta_-)}\left\vert \uparrow \downarrow \right\rangle\! \left\langle \downarrow \uparrow \right\vert \\
    &\quad-\tfrac{1}{2i} \sqrt{\kappa_a \kappa_b} \|t_{ab}^* - t_{ba}\| e^{i (\phi_a - \phi_b + \delta_-)}\left\vert \downarrow \uparrow \right\rangle\! \left\langle \uparrow \downarrow \right\vert.
\end{split}
\end{equation}
The first two terms of $H_{\ms{eff}}$ are merely energy shifts of the $2$- and $0$-excitation states and the remaining terms act only in the single-excitation subspace.  For a system initialized in the pure state $\left\vert \uparrow \downarrow \right\rangle$, the total master equation will never populate the state $\left\vert \uparrow \uparrow \right\rangle$.  However, as the total probability is conserved, the probability to be in $\left\vert \downarrow \downarrow \right\rangle$ will be unity less the total probability to be in the single-excitation subspace.  This implies that for this specific initial condition, the entire system evolution is fully characterized by the evolution in the single-excitation subspace.

In terms of the Bloch-sphere picture with $|0\rangle \equiv \left\vert \uparrow \downarrow \right\rangle$ and $|1\rangle \equiv \left\vert \downarrow \uparrow\right\rangle$, we can parameterize $\rho$ as
\begin{equation}
\begin{split}
    \rho(t) &= \tfrac{1}{2}\left[\,\mathsf{b}_0(t) I + \mathsf{b}_x(t) \sigma_x + \mathsf{b}_y(t) \sigma_y + \mathsf{b}_z(t) \sigma_z \right] \\
   &\quad + \left[1 - \mathsf{b}_0(t)\,\right]\, \left\vert \downarrow \downarrow \right\rangle\! \left\langle \downarrow \downarrow  \right\vert.
\end{split}
\end{equation}

The single-excitation part of $H_{\ms{eff}}$ defines a ``spin exchange'' operator
\begin{equation}
  J = \tfrac{1}{2}\left(\mathsf{J}_0 I + \mathsf{J}_x \sigma_x + \mathsf{J}_y \sigma_y + \mathsf{J}_z \sigma_z \right)
\end{equation}
where
\begin{equation}
  \begin{split}
    \mathsf{J}_0 &= \kappa_a \operatorname{Im} (t_{aa}) + \kappa_b \operatorname{Im}(t_{bb}) \\
    \mathsf{J}_x &= - \sqrt{\kappa_a \kappa_b}\, \beta_-\, \sin( \phi_a - \phi_b + \delta_{-} ) \\
    \mathsf{J}_y &= + \sqrt{\kappa_a \kappa_b}\, \beta_-\, \cos( \phi_a - \phi_b + \delta_{-} ) \\
    \mathsf{J}_z &= \kappa_a \operatorname{Im}(t_{aa}) - \kappa_b \operatorname{Im}(t_{bb}) + {h_a}_z - {h_b}_z.
  \end{split}
\end{equation}

Computing the expectation values $\operatorname{Tr}(\frac{d\rho}{dt} \sigma_\alpha)$ for $\sigma_\alpha \in \{I, \sigma_x, \sigma_y, \sigma_z\}$, results in the coupled equations
\begin{subequations}\label{eq:blochdot}
\begin{align}
  \tfrac{d}{dt} \mathsf{b}_0 &= - \tfrac{1}{2} \mathsf{R}_0 \mathsf{b}_0 - \tfrac{1}{2} \vec{\mathsf{R}} \cdot \vec{\mathsf{b}} \label{eq:db0dt}\\
  \tfrac{d}{dt} \vec{\mathsf{b}} &= \vec{\mathsf{J}} \times \vec{\mathsf{b}} - \tfrac{1}{2} \mathsf{R}_0 \vec{\mathsf{b}} - \tfrac{1}{2} \mathsf{b}_0 \vec{\mathsf{R}}.
\end{align}
\end{subequations}
These equations can be decoupled in a particularly relevant special case.  A straightforward calculation that shows,
\begin{equation}
  \tfrac{d}{dt}  \|\vec{\mathsf{b}}\| = - \tfrac{1}{2} \mathsf{R}_0 \| \vec{\mathsf{b}}\| - \tfrac{1}{2}  \mathsf{b}_0 \vec{\mathsf{R}}\cdot \vec{e}_{\mathsf{b}}.
\end{equation}
when combined with Eq.(\ref{eq:db0dt}) gives
\begin{equation}
  \tfrac{d}{dt} \left( \|\vec{\mathsf{b}}\| -  \mathsf{b}_0 \right) ^2 = - \left( \mathsf{R}_0 - \vec{\mathsf{R}}\cdot \vec{e}_{\mathsf{b}} \right) \left( \|\vec{\mathsf{b}}\| -  \mathsf{b}_0 \right)^2.
\end{equation}
Thus if $\|\vec{\mathsf{b}}(0)\| \ne \mathsf{b}_0(0)$, they will converge exponentially in time.  More importantly, if they are equal at $t = 0$ then they will remain equal for all $t \ge 0$.  In this case, Eq.(\ref{eq:db0dt}) has the explicit solution:
\begin{equation}\label{eq:b0solution}
  \mathsf{b}_0(t) = \mathsf{b}_0(0) \exp \left[ - \frac{1}{2} \int_{0}^{t}ds\, \left( \mathsf{R}_0(s) + \vec{\mathsf{R}}(s) \cdot \vec{e}_{\mathsf{b}}(s)\right) \right].
\end{equation}
For the pure-state initial condition, $\rho(0) = \left\vert \uparrow \downarrow \right\rangle \! \left\langle \uparrow \downarrow \right\vert$, we have $\|\vec{\mathsf{b}}(0)\| = \mathsf{b}_0(0) = 1$.  A final exercise in vector calculus shows that for the pure-state initial condition, $\vec{e}_{\mathsf{b}}$ has the equation of motion
\begin{equation}\label{eq:eBdot}
  \tfrac{d}{dt} \vec{e}_{\mathsf{b}}(t) = \vec{\mathsf{J}} \times \vec{e}_{\mathsf{b}}(t)  - \tfrac{1}{2} \left(  \vec{\mathsf{R}}  - \vec{\mathsf{R}}\cdot \vec{e}_{\mathsf{b}}(t)\, \vec{e}_{\mathsf{b}}(t) \right).
\end{equation}

\subsection{Dark-state controls}
\label{ss:dark}

Eq.(\ref{eq:b0solution}) shows that if the Bloch vector points in the opposite direction from $\vec{R}$, i.e.\ $\vec{R}/\|\vec{R}\| \equiv \vec{e}_{\mathsf{R}} = - \vec{e}_{\mathsf{b}}$, then the probability of loosing the single system excitation is minimized.  This leads to the inequality,
\begin{equation}
\begin{split}
  \mathsf{b}_0(t) &\le \exp \left[ - \frac{1}{2} \int_{0}^{t}ds\, \left( \mathsf{R}_0(s) - \|\vec{\mathsf{R}}(s)\| \right) \right] \\
   &= \exp \left[ - \int_{0}^{t}ds\, \Gamma_{d}(s) \right].
\end{split}
\end{equation}
In other words, we again see that subradiant decay rate bounds the degree of radiant loss.

We have already shown that if $0 < \kappa_a(0) \ll \kappa_b(0)$ then $\vec{R}(0)/\|\vec{R}(0)\| \equiv \vec{e}_{\mathsf{R}}(0) \approx -\vec{e}_{\mathsf{b}}(0)$. Thus our control objective is to perform a $\pi$-rotation pulse for the bloch vector $\vec{\mathsf{b}}(t)$ while maintaining the relation $\vec{e}_{\mathsf{b}}(t) = - \vec{e}_{\mathsf{R}}(t)$ throughout.  As $\vec{e}_{\mathsf{b}}(t)$ is the solution to the first-order ODE given in Eq.(\ref{eq:eBdot}), the evolution will remain in the dark state so long as $ d\vec{e}_{\mathsf{b}}(t)/dt = - d\vec{e}_{\mathsf{R}}(t)/dt$.  Note that when $\vec{e}_{\mathsf{b}} = \pm \vec{e}_{\mathsf{R}}$  the second term in Eq.(\ref{eq:eBdot}) is zero and only the coherent rotation caused by $\vec{\mathsf{J}}$ is relevant.  Thus evaluating Eq.(\ref{eq:eBdot}) at $\vec{e}_{\mathsf{b}} = - \vec{e}_{\mathsf{R}}$ shows that the derivative requirement leads to the constraint,
\begin{equation}\label{eq:darkConstraint}
  \tfrac{d}{dt}\vec{e}_{\mathsf{R}}(t) = \vec{\mathsf{J}} \times \vec{e}_{\mathsf{R}}.
\end{equation}
If $\vec{e}_{\mathsf{R}}(t)$ satisfies this constraint and $0 < \kappa_a(0) \ll \kappa_b(0)$, then $\vec{e}_{\mathsf{b}}$ will faithfully track the dark state.

Eq.(\ref{eq:darkConstraint}) can be satisfied in a number of different ways.  Here we derive a relatively simple solution, where the sender only needs to switch on and off the decay rate $\kappa_a$ so that it is equal to some nonzero constant value $\kappa_0$ for a pre-specified total time $T$.  The receiver then simultaneously varies the parameters $\kappa_b(t)$ and ${h_b}_z(t)$ with precalculated wave forms. We note that the solution for the ideal case is already known and has an analytic form~\cite{jahne_high-fidelity_2007}.  All other parameters $\phi_a$, $\phi_b$, ${h_a}_z$, $t_{ij}$, etc. are assumed to be known constants.  With no further loss of generality, we choose a phase reference such that $\phi_a - \phi_b = -\delta_+$, thereby setting $R_y = 0$. Any other choice for this phase difference corresponds merely to a fixed rotation of the Bloch ball about the z-axis.

Given the above choice of phase, a simple calculation shows that
\begin{equation}\label{eq:JcrossR}
  \vec{\mathsf{J}} \times \vec{\mathsf{R}} = \mathsf{J}_y \mathsf{R}_z\, \vec{e}_x  + \left(\mathsf{J}_z \mathsf{R}_x - \mathsf{J}_x \mathsf{R}_z \right) \vec{e}_y - \mathsf{J}_y \mathsf{R}_x\, \vec{e}_z.
\end{equation}
However as $\mathsf{R}_y = 0$, it must be the case that $\vec{e}_y \cdot \frac{d}{dt} \vec{e}_{\mathsf{R}}(t) = 0$. Thus in order for to satisfy Eq.(\ref{eq:darkConstraint}) it must be true that
\begin{equation}
  \mathsf{J}_z \mathsf{R}_x - \mathsf{J}_x \mathsf{R}_z = 0.
\end{equation}
Other than the trivial solution in which either $\kappa_a$ or $\kappa_b$ is zero, we must have
\begin{equation}\label{eq:eYconstraint}
  \frac{\mathsf{J}_z}{\mathsf{R}_z} = \frac{\mathsf{J}_x}{\mathsf{R}_x} = \frac{\beta_-}{2 \beta_+} \sin(\delta_+ - \delta_- ),
\end{equation}
which is constant.  Solving Eq.(\ref{eq:eYconstraint}) for ${h_b}_z(t)$ in terms of $\kappa_b$ we obtain
\begin{equation}
  \begin{split}
    {h_b}_z(t) & = \kappa_b(t)\, \left[\eta_b \frac{\beta_-}{2 \beta_+} \sin(\delta_+ - \delta_- ) - \operatorname{Im}(t_{bb})\right] \\
    & \quad - \kappa_0 \,\left[ \eta_a \frac{\beta_-}{2 \beta_+} \sin(\delta_+ - \delta_- ) - \operatorname{Im}(t_{aa}) \right]\\
    &\quad + {h_a}_z.
  \end{split}
\end{equation}
Returning now to the LHS of Eq.(\ref{eq:darkConstraint}), an exercise in vector calculus shows that
\begin{equation}\label{eq:eRdot}
\begin{split}
    \frac{d}{dt}\vec{e}_{\mathsf{R}}(t) &= \frac{1}{\|\vec{\mathsf{R}}\|} \left( \frac{d}{dt} \vec{\mathsf{R}} - \vec{e}_{\mathsf{R}}(t)\, \vec{e}_{\mathsf{R}}(t) \cdot \frac{d}{dt} \vec{\mathsf{R}} \right) \\
    & = \frac{1}{\|\vec{\mathsf{R}}\|} \left[ \left(\frac{\mathsf{R}_z}{\|\vec{\mathsf{R}}\|} \right)^2 \frac{d}{dt}\mathsf{R}_x  - \frac{\mathsf{R}_z \mathsf{R}_x}{\|\vec{\mathsf{R}}\|^2} \frac{d}{dt}\mathsf{R}_z \right] \vec{e}_x \\
    & \ + \frac{1}{\|\vec{\mathsf{R}}\|} \left[ \left(\frac{\mathsf{R}_x}{\|\vec{\mathsf{R}}\|} \right)^2 \frac{d}{dt}\mathsf{R}_z  - \frac{\mathsf{R}_z \mathsf{R}_x}{\|\vec{\mathsf{R}}\|^2} \frac{d}{dt}\mathsf{R}_x \right] \vec{e}_z.
\end{split}
\end{equation}
Taking the $\vec{e}_{z}$ of Eq.(\ref{eq:eRdot}), multiplying by $\|\vec{R}\|$, and setting it equal to $\vec{\mathsf{J}}\times \vec{\mathsf{R}} \cdot \vec{e}_z$ gives us the requirement that
\begin{equation}
  \frac{\mathsf{R}_x^2}{\|\vec{\mathsf{R}}\|^2} \frac{d}{dt}\mathsf{R}_z  - \frac{\mathsf{R}_x \mathsf{R}_z}{\|\vec{\mathsf{R}}\|^2} \frac{d}{dt}\mathsf{R}_x = - \mathsf{J}_y R_x.
\end{equation}
In order of this equality to hold for $\mathsf{R}_x \ne 0$, it must be the case that
\begin{equation}
\begin{split}
   \frac{\mathsf{R}_z}{\mathsf{R}_x} \frac{d}{dt}\mathsf{R}_x - \frac{d}{dt}\mathsf{R}_z & =\frac{\mathsf{J}_y}{\mathsf{R}_x} \|\vec{\mathsf{R}}\|^2 \\
   & = \cos(\delta_+ - \delta_-)\, \frac{\beta_-}{2\beta_+}\, \|\vec{\mathsf{R}}\|^2 .
\end{split}
\end{equation}
Using the basic definitions of $\vec{\mathsf{R}}$ and $\mathsf{R}_0$ from Eq.(\ref{eq:R}), the LHS of the above equation simplifies to
\begin{equation}
  \frac{\mathsf{R}_z}{\mathsf{R}_x} \frac{d}{dt}\mathsf{R}_x - \frac{d}{dt}\mathsf{R}_z = \mathsf{R}_0 \frac{1}{2 \kappa_b} \frac{d}{dt}\kappa_b.
\end{equation}
Thus we finally obtain an explicit ODE that shows how to control $\kappa_b(t)$ with time in order to obtain a transfer with minimal loss:
\begin{equation}\label{eq:KbDot}
  \begin{split}
    \frac{d}{dt}\kappa_b(t) &= \cos(\delta_+ - \delta_- )\,\kappa_b \frac{\beta_-}{\beta_+} \frac{\|\vec{\mathsf{R}}\|^2}{\mathsf{R}_0}.
  \end{split}
\end{equation}

For a perfect unidirectional channel we have already seen that $\|\vec{\mathsf{R}}\| = \mathsf{R}_0$ because $\Gamma_d = 0$. In this case Eq.(\ref{eq:KbDot}) takes the form $\dot{\kappa}_b = c_1 \kappa_b + c_2 \kappa_b^2$, for some constants $c_1$ and $c_2$.  This simplified equation has a known analytic solution, which reproduces the control solution obtained in reference~\cite{jahne_high-fidelity_2007}.

Note that in general, the solution of Eq.(\ref{eq:KbDot}) ensures that the joint system remains aligned with the subradiant state $|D\rangle$, which does not necessarily guarantee that the the total evolution results in a $\pi$ rotation on the Bloch sphere.  However, in order for $\vec{e}_{\mathsf{R}}(0) \approx - \vec{e}_z$, we have the initial condition of $\kappa_0 \ll \kappa_b(0)$.  If, at the terminal time $t_f$, we have $\kappa_b(t_f) \ll \kappa_0$, then $\vec{e}_{\mathsf{R}} \approx \vec{e}_z$ and the $\pi$-pulse was achieved.  This terminal condition can certainly be arranged if we have $\frac{d}{dt}\kappa_b(t) < 0$ for all $t \le T$.

By definition $\kappa_0$, $\beta_{\pm}$, and $\| \vec{\mathsf{R}} \|^2$ are all nonnegative.  So long as $\eta_a$ and $\eta_b$ are both positive, $\mathsf{R}_0$ is also nonnegative.  (This is always the case for weak retro-refections).  Thus if $\cos(\delta_+ - \delta_-) < 0$, then $\frac{d}{dt}\vec{e}_{\mathsf{R}}(t) \le 0$ for all $t$.  For a perfect unidirectional channel, $\cos(\delta_+ - \delta_-) = -1$.  However, if $\cos(\delta_+ - \delta_-) >0$, we can obtain a solution simply by reversing the roles of sender and receiver.  Thus $\cos(\delta_+ - \delta_-)$ serves as a measure of the networks nonreciprocity.

\subsection{Numerical simulations}
\label{ss:sims}

\begin{figure*}[tbh]
    \begin{center}
    \includegraphics[width=7in]{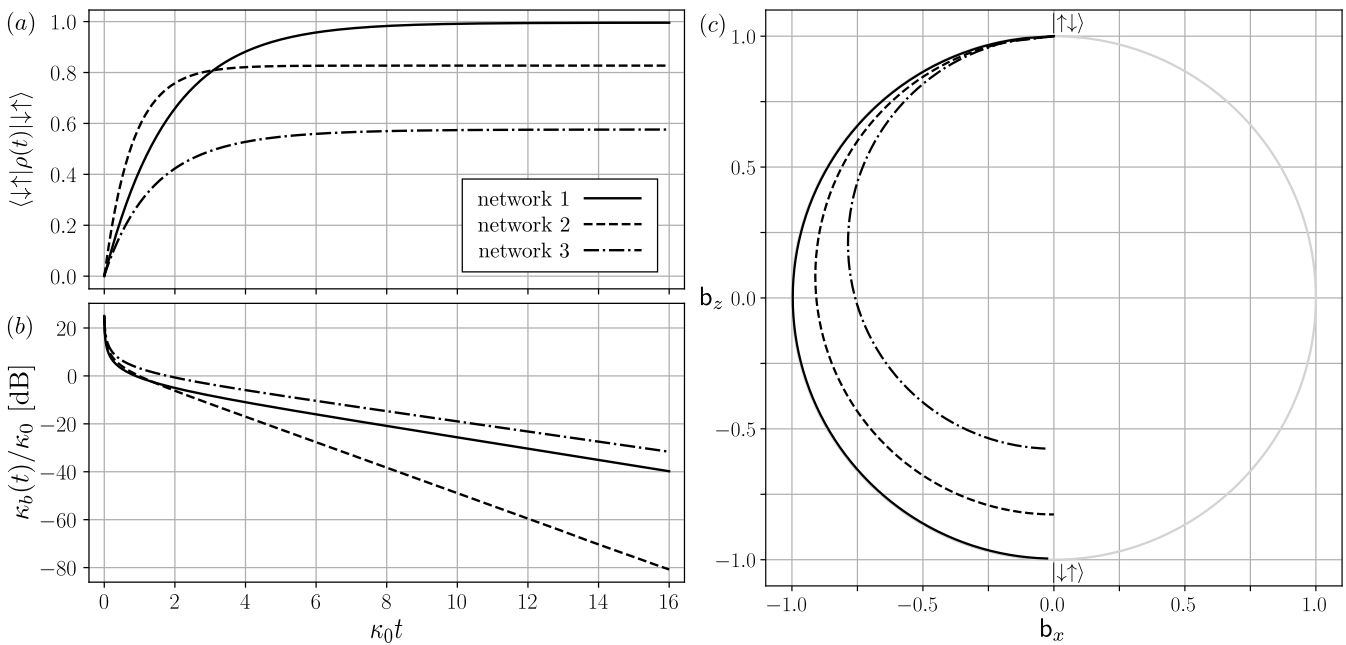}
        \caption{\label{fig:DarkSimulations} Dark state evolution.  Numerical simulations for 3 different network parameters, see text.  (a) $\left\vert \uparrow \downarrow \right\rangle \mapsto \left\vert \uparrow \downarrow \right\rangle$ transfer probability vs. time.  (b) Receiver coupling rate $\kappa_b$, relative to a fixed $\kappa_a = \kappa_0$ in dB.  (c) Bloch vector trajectories, $\vec{\mathsf{b}}(t)$, shown in the $x$-$z$ plane of the Bloch ball.}
	\end{center}
\end{figure*}

In Fig.~\ref{fig:DarkSimulations} we show the results of numerical simulations of the dark state evolution, namely Eqs.(\ref{eq:blochdot}) and (\ref{eq:KbDot}), for 3 different networks.  Each configuration is an imperfect instance of the circulator network of Fig.~\ref{fig:reductions}$(c)$.  The imperfections are introduced in two ways.  First, the propagation phase introduced by the bidirectional connection, via $\mathbf{W}$, between the two circulators is varied. Second, the ideal block-diagonal $\mathbf{S}$ matrix is replaced by a similarly block diagonal random unitary, where each sub-matrix is constrained to be within a certain distance of the identity.

Fig.~\ref{fig:DarkSimulations}$(a)$ displays the overlap of the evolving state with the target state $\left\vert \downarrow \uparrow \right \rangle$, i.e.\ the probability of a successful transfer as a function of time.  Fig.~\ref{fig:DarkSimulations}$(b)$ shows how $\kappa_b(t)$ is varied with time to achieve the transfer, relative to the value of $\kappa_0$ in dB, with an initial value of $\kappa_b(0)/ \kappa_0 = 25$ dB.  With our particular choice of phase, the evolution of $\vec{\mathsf{b}}(t)$ is constrained to the $x$-$z$ plane of the Bloch ball. In Fig.~\ref{fig:DarkSimulations}$(c)$ we plot the trajectory made by $\vec{\mathsf{b}}(t)$ in this plane for each of the networks.

Network configuration $1$ demonstrates that near-perfect state transfer is possible, even when the network is significantly far from the ideal unidirectional connection.  The imperfect circulators have retro-reflections in the range $0.04 \lesssim |r_{ii}|^2 \lesssim 0.15$, with a total $b \mapsto a$ transfer coefficient with magnitude $|t_{ab}| = 0.02$.  In spite of this, the coherent effects collude in such a way as to ensure that the criteria of Eq.(\ref{eq:perfectConstraint}) is nearly satisfied, with $\eta_a \eta_b - \beta_+^2 = 0.001$.  At the terminal time, the probability for the system to be measured in $\left \vert \downarrow \uparrow \right \rangle$ is $0.996$.

The performance of networks with randomized imperfections with magnitudes similar to network 1 varies considerably, with the excellent performance of network 1 being atypical. Network $2$, for which the range chosen for the imperfections is $0.03 \lesssim |r_{ii}|^2 \lesssim 0.14$ (similar to those of network 1) gives a typical example of the resulting performance. Despite the fact that Eq.(\ref{eq:perfectConstraint}) is far from satisfied for network 2, $\eta_a \eta_b - \beta_+^2 = 0.147$, the terminal probability for a successful transfer is $0.827$.

Network $3$ shows a particularly adverse case, with $0.42 \lesssim |r_{ii}|^2 \lesssim 0.84$.  Additionally the asymmetry in the $a \mapsto b$ transfer is particularly impeded, with $\cos(\delta_+ - \delta_-) = -0.151$.  unsurprisingly, the final success probability is only $0.576$, although poorer performance can occur for similar parameters.

Fig.~\ref{fig:DarkSimulations}$(b)$ show all three optimal protocols (solutions of Eq.(\ref{eq:KbDot})) for $\kappa_b(t)$, all of which involve a rapid descent from the regime where $\kappa_b \gg \kappa_a$ and most of the transfer time spent in the asymptotic limit where $\kappa_b \ll \kappa_a$. This suggests that when considering practical limitations to the control resources, varying $\kappa_a$ in addition to $\kappa_b$ will be helpful and may be necessary. When $\partial_t \kappa_a \ne 0$,  Eq.(\ref{eq:KbDot}) remains pertinent, but as an equation of motion for the ratio $\kappa_b/\kappa_a$ in terms of the rescaled time, $\partial_t \mapsto \partial_\tau$ where $\tau(t) \equiv \int_0^t \kappa_a(s)\, ds$.  In this light, making $\kappa_a$ time dependent results in the compression/expansion of the $\kappa_0 t$ axis.

\section{Summary and outlook} \label{sec:summary}

We have elucidated how and when the network contraction theory of Gough and James can be applied to physical networks of input-output systems. This theory allows one to accurately model networks containing weak loops that cause the fields to circulate in the network. We have shown that, in particular, the method provides the first analytically tractable way to handle retro-reflections that are a common and important source of imperfection in superconducting and photonic circuits. We have presented a formulation of the method that requires only a single matrix inversion, and is thus efficient for analytical calculations. We have also re-derived the theory in the language typically used by physicists, making it easily accessible. We have provided an explicit example in which we apply the method to the problem of transmitting entanglement between two qubits connected via two imperfect circulators. This example showed that despite the retro-reflections it was possible to obtain a largely analytic solution to the problem of maximizing the probability of a successful transfer.

Networks with weak loops can be thought of as quantum feedback networks. The fact that the effective input-output description of a network with weak loops can be obtained using a single matrix inversion may well have use in establishing systematic methods for the design of quantum feedback networks. Given an effective input-output network that one wishes to construct, the network topology that would induce this effective dynamics can thus also be obtained by inverting a matrix. This does not by itself solve the network design problem, since there is no guarantee that all effective input-output models can be obtained by constructing loopy networks under a given set of constraints. Nevertheless, the question of what input-output dynamics can be engineered via the introduction of feedback loops, under experimentally motivated constraints, is a interesting question for future work, and one for which the technique presented here may well be useful.

\begin{acknowledgments}
The research reported here was sponsored by the Army Research Laboratory and was accomplished under Cooperative Agreement Number W911NF-16-2-0170. The views and conclusions contained in this document are those of the authors and should not be interpreted as representing the official policies, either expressed or implied, of the Army Research Laboratory or the U.S. Government. The U.S. Government is authorized to reproduce and distribute reprints for Government purposes notwithstanding any copyright notation herein.

This research was supported in part by an appointment to the Postgraduate Research Participation Program at the U.S. Army Research Laboratory administered by the Oak Ridge Institute for Science and Education through an interagency agreement between the U.S. Department of Energy and USARL.
\end{acknowledgments}

\appendix

\section{Algebraic derivation of the effective model \label{app:algebra}}

Here we present the algebraic derivation showing that merely by  enforcing the constraint,
\begin{equation}\label{eq:app:Constraint}
\begin{split}
  \mathbf{b}_{\ms{in}}^{\ms{int}} = \mathbf{Wb}_{\ms{out}}^{\ms{int}} ,
\end{split}
\end{equation}
which can also be written as $\mathbf{I}_{\ms{i}} \mathbf{b}_{\ms{in}} = \mathbf{Wb}_{\ms{out}}$, an unconnected set of network elements described by the set of quantities ($\mathbf{S}$, $\mathbf{L}$, $\mathbf{H}$) can be described by an effective input-output model given by the set of quantities  ($\mathbf{S}_{\ms{eff}}$,$\mathbf{L}_{\ms{eff}}$,$H_{\ms{eff}}$).  For simplicity we set $H_{\ms{sys}} = 0$ as it will play no role.  For reference, the equations of the input-output formalism that describe the unconnected network elements are
\begin{equation}\label{eq:app:IOtotal}
  \mathbf{b}_{\ms{out}} = \mathbf{Sb}_{\ms{in}} + \mathbf{L},
\end{equation}
and
\begin{equation}\label{eq:app:EOM}
\begin{split}
  \dot{A} &= - \frac{1}{2} \left(  [A, \mathbf{L}^\dag] \mathbf{L} - \mathbf{L}^\dag [A,\mathbf{L}] \right)\\
  & \quad - [A,\mathbf{L}^\dag ] \mathbf{S} \mathbf{b}_{\ms{in}} + \left( \mathbf{Sb}_{\ms{in}}\right)^\dag [A, \mathbf{L} ],
\end{split}
\end{equation}
and those of the resulting effective input-output description are
\begin{equation}\label{eq:app:IOeff}
  \mathbf{b}_{\ms{out}}^{\ms{ext}} = \mathbf{S}_{\ms{eff}}\mathbf{b}_{\ms{in}}^{\ms{ext}} + \mathbf{L}_{\ms{eff}}
\end{equation}
and
\begin{equation}\label{eq:app:EOMeff}
\begin{split}
  \dot{A} &= + \tfrac{i}{\hbar}\left[H_{\ms{eff}}, A \right]  - \tfrac{1}{2} \left(  [A, \mathbf{L}^\dag_{\ms{eff}}] \mathbf{L}_{\ms{eff}} - \mathbf{L}^\dag_{\ms{eff}} [A,\mathbf{L}_{\ms{eff}}] \right)\\
  & \quad - [A,\mathbf{L}^\dag_{\ms{eff}} ] \mathbf{S}_{\ms{eff}} \mathbf{b}^{\ms{ext}}_{\ms{in}} + \left( \mathbf{S}_{\ms{eff}}\mathbf{b}^{\ms{ext}}_{\ms{in}}\right)^\dag [A, \mathbf{L}_{\ms{eff}} ].
\end{split}
\end{equation}

\subsection{The external input-output relation}
Starting from Eq.(\ref{eq:app:IOtotal}), we first decompose $\mathbf{b}_{\ms{in}}$ into its internal and external components,
\begin{equation}
\mathbf{b}_{\ms{out}} = \mathbf{S}\left( \mathbf{X}_{\ms{i}}\mathbf{b}_{\ms{in}} + \mathbf{I}_{\ms{i}}\mathbf{b}_{\ms{in}} \right) + \mathbf{L}.
\end{equation}
substituting the constraint as written in the second line of Eq.(\ref{eq:app:Constraint}) shows,
\begin{equation}
\mathbf{b}_{\ms{out}} = \mathbf{S} \mathbf{X}_{\ms{i}}\mathbf{b}_{\ms{in}} + \mathbf{SW} \mathbf{b}_{\ms{out}} + \mathbf{L}.
\end{equation}
Subtracting $\mathbf{SW} \mathbf{b}_{\ms{out}}$ from both sides results in,
\begin{equation}
\left(\ident - \mathbf{SW}\right)\mathbf{b}_{\ms{out}} = \mathbf{S} \mathbf{X}_{\ms{i}}\mathbf{b}_{\ms{in}} + \mathbf{L}.
\end{equation}
Wherever $\ident - \mathbf{SW}$ is invertible, or equivalently, when the series $\sum_{n} (\mathbf{SW})^n$ converges we have,
\begin{equation}\label{eq:app:Boutsubs}
\mathbf{b}_{\ms{out}} = \frac{1}{\ident - \mathbf{SW}}\mathbf{S} \mathbf{X}_{\ms{i}}\mathbf{b}_{\ms{in}} + \frac{1}{\ident - \mathbf{SW}} \mathbf{L}.
\end{equation}
Thus projecting onto the external outputs, shows that
\begin{equation}
  \mathbf{b}_{\ms{out}}^{\ms{ext}} = \mathbf{S}_{\ms{eff}} \mathbf{b}_{\ms{in}}^{\ms{ext}} + \mathbf{L}_{\ms{eff}}
\end{equation}
where
\begin{equation}\label{eq:app:Seff}
  \mathbf{S}_{\ms{eff}} = \mathbf{X}_{\ms{o}} \frac{1}{\ident - \mathbf{SW}} \mathbf{S} \mathbf{X}_{\ms{i}},
\end{equation}
and
\begin{equation}\label{eq:app:Leff}
  \mathbf{L}_{\ms{eff}} = \mathbf{X}_{\ms{o}} \frac{1}{\ident - \mathbf{SW}} \mathbf{L}.
\end{equation}

\subsection{The Heisenberg-Langevin equation of motion}
Here we show that expressing Eq.(\ref{eq:app:EOM}) in terms of $\mathbf{S}_{\ms{eff}}$ and $\mathbf{L}_{\ms{eff}}$ results in an additional term in $H_{\ms{eff}}$.  The first line of attack is to write $\mathbf{Sb}_{\ms{in}}$ in terms of the external inputs and system sources.  In other words,
\begin{equation}
\begin{split}
  \mathbf{Sb}_{\ms{in}} &= \mathbf{S X}_{\ms{i}}\mathbf{b}_{\ms{in}} + \mathbf{SI}_{\ms{i}}\mathbf{b}_{\ms{in}} \\
    &= \mathbf{S X}_{\ms{i}}\mathbf{b}_{\ms{in}} + \mathbf{SW}\mathbf{b}_{\ms{out}}\\
    &= \mathbf{S X}_{\ms{i}}\mathbf{b}_{\ms{in}} + \mathbf{SWSb}_{\ms{in}} + \mathbf{SWL}.
\end{split}
\end{equation}
Collecting all terms involving $\mathbf{Sb}_{\ms{in}}$ on the left hand side and then acting on both sides with $(\ident - \mathbf{SW})^{-1}$ shows that
\begin{equation}\label{eq:app:Sbin}
  \mathbf{Sb}_{\ms{in}} = \frac{1}{\ident - \mathbf{SW}}\mathbf{S X}_{\ms{i}}\mathbf{b}_{\ms{in}} + \frac{\mathbf{SW}}{\ident - \mathbf{SW}}\mathbf{L}.
\end{equation}
Substituting this into Eq.(\ref{eq:app:EOM}) and collecting like commutator terms gives us
\begin{widetext}
\begin{equation}\label{eq:app:EOM+Constrant}
\begin{split}
    \dot{A} &= - [A, \mathbf{L}^\dag] \left(\frac{1}{2}\ident + \frac{\mathbf{SW}}{\ident - \mathbf{SW}} \right) \mathbf{L} + \mathbf{L}^\dag\left( \frac{1}{2}\ident + \frac{(\mathbf{SW})^\dag}{\ident - (\mathbf{SW})^\dag} \right) [A,\mathbf{L}] \\
     & \quad - [A,\mathbf{L}^\dag ] \frac{1}{\ident - \mathbf{SW}}\mathbf{S X}_{\ms{i}}\mathbf{b}_{\ms{in}} + \left( \frac{1}{\ident - \mathbf{SW}}\mathbf{S X}_{\ms{i}}\mathbf{b}_{\ms{in}} \right)^\dag [A, \mathbf{L} ].
\end{split}
\end{equation}
\end{widetext}
Before working through further simplifications, it is useful to identify some expected terms.  The effective equation of motion Eq.(\ref{eq:app:EOMeff}) contains the terms $A \mathbf{L}^\dag_{\ms{eff}}\mathbf{S}_{\ms{eff}}\mathbf{b}^{\ms{ext}}_{\ms{in}}$ and $\mathbf{L}^\dag_{\ms{eff}}A\mathbf{S}_{\ms{eff}}\mathbf{b}^{\ms{ext}}_{\ms{in}}$.  However, multiplying Eq.(\ref{eq:app:Seff}) by the adjoint of Eq.(\ref{eq:app:Leff}) shows that
\begin{equation}
  \mathbf{L}^\dag_{\ms{eff}} \mathbf{S}_{\ms{eff}}  = \mathbf{L}^\dag \frac{1}{\ident - (\mathbf{SW})^\dag} \mathbf{X}_{\ms{o}}^2 \frac{1}{\ident - \mathbf{SW}} \mathbf{SX}_{\ms{i}}.
\end{equation}
This expression can be simplified by first noting that $\mathbf{X}_{\ms{o}}^2 = \mathbf{X}_{\ms{o}}$, and second by writing
\begin{equation}
\begin{split}
  \mathbf{X}_{\ms{o}} &= \ident - \mathbf{W}^\dag \mathbf{W} = \ident - \mathbf{W}^\dag \mathbf{S}^\dag\mathbf{SW} \\
    &= \ident - (\mathbf{SW})^\dag + \ident - \mathbf{SW} - [\ident - (\mathbf{SW})^\dag][\ident - \mathbf{SW}].
\end{split}
\end{equation}
To obtain the second equality we used the fact that $\mathbf{S}$ is unitary, and the third equality, while true, is useful only in hindsight. However, this rather opaque rewriting leads to the relation
\begin{equation}\label{eq:app:trick}
\begin{split}
  \tfrac{1}{\ident - (\mathbf{SW})^\dag} \mathbf{X}_{\ms{o}} \tfrac{1}{\ident - \mathbf{SW}} &= \frac{1}{\ident - \mathbf{SW}} + \frac{1}{\ident - (\mathbf{SW})^\dag} - \ident \\
   &=\frac{1}{\ident - \mathbf{SW}} +  \frac{(\mathbf{SW})^\dag}{\ident - (\mathbf{SW})^\dag}.
\end{split}
\end{equation}
This is particularly useful as it shows that
\begin{equation}
\begin{split}
  A \mathbf{L}^\dag_{\ms{eff}} \mathbf{S}_{\ms{eff}} &= A \mathbf{L}^\dag \left( \frac{1}{\ident - \mathbf{SW}} +  \frac{(\mathbf{SW})^\dag}{\ident - (\mathbf{SW})^\dag} \right) \mathbf{SX}_{\ms{i}} \\
  &= A \mathbf{L}^\dag \frac{1}{\ident - \mathbf{SW}}\mathbf{SX}_{\ms{i}} + A \mathbf{L}^\dag\frac{1}{\ident - (\mathbf{SW})^\dag}\mathbf{W}^\dag \mathbf{X}_{\ms{i}} \\
  &= A \mathbf{L}^\dag \frac{1}{\ident - \mathbf{SW}}\mathbf{SX}_{\ms{i}},
\end{split}
\end{equation}
where the second term is ultimately zero because $\mathbf{S}$ is unitary and $\mathbf{W}^\dag$ is orthogonal to $\mathbf{X}_{\ms{i}}$. Furthermore, because we have assumed that $[\mathbf{S},A] = [\mathbf{SW}, A] = 0$, we also find that
\begin{widetext}
\begin{equation}
\begin{split}
  \mathbf{L}^\dag_{\ms{eff}}A \mathbf{S}_{\ms{eff}} &= \mathbf{L}^\dag A \frac{1}{\ident - (\mathbf{SW})^\dag} \mathbf{X}_{\ms{o}}^2 \frac{1}{\ident - \mathbf{SW}} \mathbf{SX}_{\ms{i}} = \mathbf{L}^\dag A \frac{1}{\ident - \mathbf{SW}}\mathbf{SX}_{\ms{i}}.
\end{split}
\end{equation}
Combining the previous two relations gives
\begin{equation}
  [A, \mathbf{L}_{\ms{eff}}^\dag] \mathbf{S}_{\ms{eff}} \mathbf{b}_{\ms{in}}^{\ms{ext}} = [A, \mathbf{L}^\dag] \frac{1}{\ident - \mathbf{SW}}\mathbf{SX}_{\ms{i}} \mathbf{b}_{\ms{in}} ,
\end{equation}
and thus we conclude that the second line of Eq.(\ref{eq:app:EOM+Constrant}) is indeed equal to the second line of Eq.(\ref{eq:app:EOMeff}).

To show that the first line of Eq.(\ref{eq:app:EOMeff}) also follows from Eq.(\ref{eq:app:EOM+Constrant}), consider the parenthetical expression in the first term.  The trick of Eq.(\ref{eq:app:trick}) does not immediately apply to this expression.  However, it is true that
\begin{equation} \label{eq:app:trick2}
    \tfrac{1}{2}\ident +  \tfrac{\mathbf{SW}}{\ident - \mathbf{SW}} = \tfrac{1}{2} \left(\ident + 2 \tfrac{\mathbf{SW}}{\ident - \mathbf{SW}} \right)
     = \tfrac{1}{2} \left( \tfrac{1}{\ident - (\mathbf{SW})^\dag}\mathbf{X}_{\ms{o}}^2 \tfrac{1}{\ident - \mathbf{SW}} + \tfrac{1}{\ident - \mathbf{SW}} - \tfrac{1}{\ident - (\mathbf{SW})^\dag}\right),
\end{equation}
which follows from Eq.(\ref{eq:app:trick}).
Now consider the full first line in Eq.(\ref{eq:app:EOM+Constrant}).   Expanding out the commutators and using the fact that $A$ commutes with any function of $\mathbf{SW}$ or its adjoint, results in
\begin{multline}
  - [A, \mathbf{L}^\dag] \left(\tfrac{1}{2}\ident + \tfrac{\mathbf{SW}}{\ident - \mathbf{SW}} \right) \mathbf{L} +\mathbf{L}^\dag\left( \tfrac{1}{2}\ident + \tfrac{(\mathbf{SW})^\dag}{\ident - (\mathbf{SW})^\dag} \right) [A,\mathbf{L}] \\
  = -\tfrac{1}{2} A \mathbf{L}^\dag \left(\tfrac{1}{\ident - (\mathbf{SW})^\dag}\mathbf{X}_{\ms{o}}^2 \tfrac{1}{\ident - \mathbf{SW}}\right) \mathbf{L}  - \tfrac{1}{2} A \mathbf{L}^\dag \left( \tfrac{1}{\ident - \mathbf{SW}} - \tfrac{1}{\ident - (\mathbf{SW})^\dag} \right) \mathbf{L} \\
  - \tfrac{1}{2} \mathbf{L}^\dag \left(\tfrac{1}{\ident - (\mathbf{SW})^\dag}\mathbf{X}_{\ms{o}}^2 \tfrac{1}{\ident - \mathbf{SW}}\right) \mathbf{L} A + \tfrac{1}{2} \mathbf{L}^\dag \left( \tfrac{1}{\ident - \mathbf{SW}} - \tfrac{1}{\ident - (\mathbf{SW})^\dag} \right) \mathbf{L} A
  + \mathbf{L}^\dag A \left(\tfrac{1}{\ident - (\mathbf{SW})^\dag}\mathbf{X}_{\ms{o}}^2 \tfrac{1}{\ident - \mathbf{SW}}\right) \mathbf{L}.
\end{multline}
By defining the effective Hamiltonian
\begin{equation}
 H_{\ms{eff}} \equiv \frac{\hbar}{2 i} \mathbf{L}^\dag \left( \tfrac{1}{\ident - \mathbf{SW}} - \tfrac{1}{\ident - (\mathbf{SW})^\dag} \right)\mathbf{L},
\end{equation}
and utilizing the definition of $\mathbf{L}_{\ms{eff}}$ shows,
\begin{equation}
\begin{split}
  - [A, \mathbf{L}^\dag] \left(\tfrac{1}{2}\ident + \tfrac{\mathbf{SW}}{\ident - \mathbf{SW}} \right) \mathbf{L}  + \mathbf{L}^\dag\left( \tfrac{1}{2}\ident + \tfrac{(\mathbf{SW})^\dag}{\ident - (\mathbf{SW})^\dag} \right) [A,\mathbf{L}] \\
  = -\tfrac{1}{2} A \mathbf{L}^\dag_{\ms{eff}} \mathbf{L}_{\ms{eff}} - \tfrac{1}{2} A \mathbf{L}^\dag_{\ms{eff}} \mathbf{L}_{\ms{eff}} + \mathbf{L}^\dag_{\ms{eff}} A \mathbf{L}_{\ms{eff}} + \tfrac{i}{\hbar}[H_{\ms{eff}}, A],
\end{split}
\end{equation}
which is equal to the first line of Eq.(\ref{eq:app:EOMeff}).

\end{widetext}

\bibliography{Photon-Receiver}

\begin{thebibliography}{38}
\expandafter\ifx\csname natexlab\endcsname\relax\def\natexlab#1{#1}\fi
\expandafter\ifx\csname bibnamefont\endcsname\relax
\def\bibnamefont#1{#1}\fi
\expandafter\ifx\csname bibfnamefont\endcsname\relax
\def\bibfnamefont#1{#1}\fi
\expandafter\ifx\csname citenamefont\endcsname\relax
\def\citenamefont#1{#1}\fi
\expandafter\ifx\csname url\endcsname\relax
\def\url#1{\texttt{#1}}\fi
\expandafter\ifx\csname urlprefix\endcsname\relax\def\urlprefix{URL }\fi
\providecommand{\bibinfo}[2]{#2}
\providecommand{\eprint}[2][]{\url{#2}}

\bibitem[{\citenamefont{Collett and Gardiner}(1984)}]{Collett84}
\bibinfo{author}{\bibfnamefont{M.~J.} \bibnamefont{Collett}} \bibnamefont{and}
\bibinfo{author}{\bibfnamefont{C.~W.} \bibnamefont{Gardiner}},
\bibinfo{journal}{Phys. Rev. A} \textbf{\bibinfo{volume}{30}},
\bibinfo{pages}{1386} (\bibinfo{year}{1984}).

\bibitem[{\citenamefont{Gardiner and Collett}(1985)}]{gardiner_input_1985}
\bibinfo{author}{\bibfnamefont{C.~W.} \bibnamefont{Gardiner}} \bibnamefont{and}
\bibinfo{author}{\bibfnamefont{M.~J.} \bibnamefont{Collett}},
\bibinfo{journal}{Phys. Rev. A} \textbf{\bibinfo{volume}{31}},
\bibinfo{pages}{3761} (\bibinfo{year}{1985}).

\bibitem[{\citenamefont{Holland et~al.}(1990)\citenamefont{Holland, Collett,
Walls, and Levenson}}]{Holland90}
\bibinfo{author}{\bibfnamefont{M.~J.} \bibnamefont{Holland}},
\bibinfo{author}{\bibfnamefont{M.~J.} \bibnamefont{Collett}},
\bibinfo{author}{\bibfnamefont{D.~F.} \bibnamefont{Walls}}, \bibnamefont{and}
\bibinfo{author}{\bibfnamefont{M.~D.} \bibnamefont{Levenson}},
\bibinfo{journal}{Phys. Rev. A} \textbf{\bibinfo{volume}{42}},
\bibinfo{pages}{2995} (\bibinfo{year}{1990}).

\bibitem[{\citenamefont{Gardiner}(1993)}]{gardiner_driving_1993}
\bibinfo{author}{\bibfnamefont{C.~W.} \bibnamefont{Gardiner}},
\bibinfo{journal}{Phys. Rev. Lett.} \textbf{\bibinfo{volume}{70}},
\bibinfo{pages}{2269} (\bibinfo{year}{1993}).

\bibitem[{\citenamefont{Kamal et~al.}(2009)\citenamefont{Kamal, Marblestone,
and Devoret}}]{Kamal09}
\bibinfo{author}{\bibfnamefont{A.}~\bibnamefont{Kamal}},
\bibinfo{author}{\bibfnamefont{A.}~\bibnamefont{Marblestone}},
\bibnamefont{and} \bibinfo{author}{\bibfnamefont{M.}~\bibnamefont{Devoret}},
\bibinfo{journal}{Phys. Rev. B} \textbf{\bibinfo{volume}{79}},
\bibinfo{pages}{184301} (\bibinfo{year}{2009}).

\bibitem[{\citenamefont{Lecocq et~al.}(2017)\citenamefont{Lecocq, Ranzani,
Peterson, Cicak, Simmonds, Teufel, and Aumentado}}]{Lecocq17}
\bibinfo{author}{\bibfnamefont{F.}~\bibnamefont{Lecocq}},
\bibinfo{author}{\bibfnamefont{L.}~\bibnamefont{Ranzani}},
\bibinfo{author}{\bibfnamefont{G.~A.} \bibnamefont{Peterson}},
\bibinfo{author}{\bibfnamefont{K.}~\bibnamefont{Cicak}},
\bibinfo{author}{\bibfnamefont{R.~W.} \bibnamefont{Simmonds}},
\bibinfo{author}{\bibfnamefont{J.~D.} \bibnamefont{Teufel}},
\bibnamefont{and}
\bibinfo{author}{\bibfnamefont{J.}~\bibnamefont{Aumentado}},
\bibinfo{journal}{Phys. Rev. Applied} \textbf{\bibinfo{volume}{7}},
\bibinfo{pages}{024028} (\bibinfo{year}{2017}).

\bibitem[{\citenamefont{Combes et~al.}(2017)\citenamefont{Combes, Kerckhoff,
and Sarovar}}]{Combes17}
\bibinfo{author}{\bibfnamefont{J.}~\bibnamefont{Combes}},
\bibinfo{author}{\bibfnamefont{J.}~\bibnamefont{Kerckhoff}},
\bibnamefont{and} \bibinfo{author}{\bibfnamefont{M.}~\bibnamefont{Sarovar}},
\bibinfo{journal}{Advances in Physics: X} \textbf{\bibinfo{volume}{2}},
\bibinfo{pages}{784} (\bibinfo{year}{2017}).

\bibitem[{\citenamefont{Gardiner and Zoller}(2004)}]{gardiner_quantum_2004}
\bibinfo{author}{\bibfnamefont{C.~W.} \bibnamefont{Gardiner}} \bibnamefont{and}
\bibinfo{author}{\bibfnamefont{P.}~\bibnamefont{Zoller}},
\emph{\bibinfo{title}{Quantum Noise}} (\bibinfo{publisher}{{Springer}},
\bibinfo{year}{2004}).

\bibitem[{\citenamefont{Jacobs}(2014)}]{jacobs_quantum_2014}
\bibinfo{author}{\bibfnamefont{K.}~\bibnamefont{Jacobs}},
\emph{\bibinfo{title}{Quantum Measurement Theory and Its Applications}}
(\bibinfo{publisher}{{Cambridge University Press}}, \bibinfo{year}{2014}).

\bibitem[{\citenamefont{Kamal et~al.}(2011)\citenamefont{Kamal, Clarke, and
Devoret}}]{Kamal11}
\bibinfo{author}{\bibfnamefont{A.}~\bibnamefont{Kamal}},
\bibinfo{author}{\bibfnamefont{J.}~\bibnamefont{Clarke}}, \bibnamefont{and}
\bibinfo{author}{\bibfnamefont{M.~H.} \bibnamefont{Devoret}},
\bibinfo{journal}{Nature Physics} \textbf{\bibinfo{volume}{7}},
\bibinfo{pages}{311} (\bibinfo{year}{2011}).

\bibitem[{\citenamefont{Naik et~al.}(2017)\citenamefont{Naik, Leung, Chakram,
Groszkowski, Lu, Earnest, McKay, Koch, and Schuster}}]{Naik17}
\bibinfo{author}{\bibfnamefont{R.~K.} \bibnamefont{Naik}},
\bibinfo{author}{\bibfnamefont{N.}~\bibnamefont{Leung}},
\bibinfo{author}{\bibfnamefont{S.}~\bibnamefont{Chakram}},
\bibinfo{author}{\bibfnamefont{P.}~\bibnamefont{Groszkowski}},
\bibinfo{author}{\bibfnamefont{Y.}~\bibnamefont{Lu}},
\bibinfo{author}{\bibfnamefont{N.}~\bibnamefont{Earnest}},
\bibinfo{author}{\bibfnamefont{D.~C.} \bibnamefont{McKay}},
\bibinfo{author}{\bibfnamefont{J.}~\bibnamefont{Koch}}, \bibnamefont{and}
\bibinfo{author}{\bibfnamefont{D.~I.} \bibnamefont{Schuster}},
\bibinfo{journal}{Nature Communications} \textbf{\bibinfo{volume}{8}},
\bibinfo{pages}{1904} (\bibinfo{year}{2017}).

\bibitem[{\citenamefont{Kandala et~al.}(2017)\citenamefont{Kandala, Mezzacapo,
Temme, Takita, Brink, Chow, and Gambetta}}]{Kandala17}
\bibinfo{author}{\bibfnamefont{A.}~\bibnamefont{Kandala}},
\bibinfo{author}{\bibfnamefont{A.}~\bibnamefont{Mezzacapo}},
\bibinfo{author}{\bibfnamefont{K.}~\bibnamefont{Temme}},
\bibinfo{author}{\bibfnamefont{M.}~\bibnamefont{Takita}},
\bibinfo{author}{\bibfnamefont{M.}~\bibnamefont{Brink}},
\bibinfo{author}{\bibfnamefont{J.~M.} \bibnamefont{Chow}}, \bibnamefont{and}
\bibinfo{author}{\bibfnamefont{J.~M.} \bibnamefont{Gambetta}},
\bibinfo{journal}{Nature} \textbf{\bibinfo{volume}{549}},
\bibinfo{pages}{242} (\bibinfo{year}{2017}).

\bibitem[{\citenamefont{Clark et~al.}(2017)\citenamefont{Clark, Lecocq,
Simmonds, Aumentado, and Teufel}}]{Clark17}
\bibinfo{author}{\bibfnamefont{J.~B.} \bibnamefont{Clark}},
\bibinfo{author}{\bibfnamefont{F.}~\bibnamefont{Lecocq}},
\bibinfo{author}{\bibfnamefont{R.~W.} \bibnamefont{Simmonds}},
\bibinfo{author}{\bibfnamefont{J.}~\bibnamefont{Aumentado}},
\bibnamefont{and} \bibinfo{author}{\bibfnamefont{J.~D.}
\bibnamefont{Teufel}}, \bibinfo{journal}{Nature}
\textbf{\bibinfo{volume}{541}}, \bibinfo{pages}{191 EP }
(\bibinfo{year}{2017}).

\bibitem[{\citenamefont{Fitzpatrick et~al.}(2017)\citenamefont{Fitzpatrick,
Sundaresan, Li, Koch, and Houck}}]{Fitzpatrick16}
\bibinfo{author}{\bibfnamefont{M.}~\bibnamefont{Fitzpatrick}},
\bibinfo{author}{\bibfnamefont{N.~M.} \bibnamefont{Sundaresan}},
\bibinfo{author}{\bibfnamefont{A.~C.~Y.} \bibnamefont{Li}},
\bibinfo{author}{\bibfnamefont{J.}~\bibnamefont{Koch}}, \bibnamefont{and}
\bibinfo{author}{\bibfnamefont{A.~A.} \bibnamefont{Houck}},
\bibinfo{journal}{Phys. Rev. X} \textbf{\bibinfo{volume}{7}},
\bibinfo{pages}{011016} (\bibinfo{year}{2017}).

\bibitem[{\citenamefont{Kelly et~al.}(2015)\citenamefont{Kelly, Barends,
Fowler, Megrant, Jeffrey, White, Sank, Mutus, Campbell, Chen
et~al.}}]{Kelly15}
\bibinfo{author}{\bibfnamefont{J.}~\bibnamefont{Kelly}},
\bibinfo{author}{\bibfnamefont{R.}~\bibnamefont{Barends}},
\bibinfo{author}{\bibfnamefont{A.~G.} \bibnamefont{Fowler}},
\bibinfo{author}{\bibfnamefont{A.}~\bibnamefont{Megrant}},
\bibinfo{author}{\bibfnamefont{E.}~\bibnamefont{Jeffrey}},
\bibinfo{author}{\bibfnamefont{T.~C.} \bibnamefont{White}},
\bibinfo{author}{\bibfnamefont{D.}~\bibnamefont{Sank}},
\bibinfo{author}{\bibfnamefont{J.~Y.} \bibnamefont{Mutus}},
\bibinfo{author}{\bibfnamefont{B.}~\bibnamefont{Campbell}},
\bibinfo{author}{\bibfnamefont{Y.}~\bibnamefont{Chen}}, \bibnamefont{et~al.},
\bibinfo{journal}{Nature} \textbf{\bibinfo{volume}{519}}, \bibinfo{pages}{66
EP } (\bibinfo{year}{2015}).

\bibitem[{\citenamefont{Yoshie et~al.}(2004)\citenamefont{Yoshie, Scherer,
Hendrickson, Khitrova, Gibbs, Rupper, Ell, Shchekin, and Deppe}}]{Yoshie04}
\bibinfo{author}{\bibfnamefont{T.}~\bibnamefont{Yoshie}},
\bibinfo{author}{\bibfnamefont{A.}~\bibnamefont{Scherer}},
\bibinfo{author}{\bibfnamefont{J.}~\bibnamefont{Hendrickson}},
\bibinfo{author}{\bibfnamefont{G.}~\bibnamefont{Khitrova}},
\bibinfo{author}{\bibfnamefont{H.~M.} \bibnamefont{Gibbs}},
\bibinfo{author}{\bibfnamefont{G.}~\bibnamefont{Rupper}},
\bibinfo{author}{\bibfnamefont{C.}~\bibnamefont{Ell}},
\bibinfo{author}{\bibfnamefont{O.~B.} \bibnamefont{Shchekin}},
\bibnamefont{and} \bibinfo{author}{\bibfnamefont{D.~G.} \bibnamefont{Deppe}},
\bibinfo{journal}{Nature} \textbf{\bibinfo{volume}{432}}, \bibinfo{pages}{200
EP } (\bibinfo{year}{2004}).

\bibitem[{\citenamefont{Reithmaier et~al.}(2004)\citenamefont{Reithmaier, Sek,
L{\"o}ffler, Hofmann, Kuhn, Reitzenstein, Keldysh, Kulakovskii, Reinecke, and
Forchel}}]{Reithmaier04}
\bibinfo{author}{\bibfnamefont{J.~P.} \bibnamefont{Reithmaier}},
\bibinfo{author}{\bibfnamefont{G.}~\bibnamefont{Sek}},
\bibinfo{author}{\bibfnamefont{A.}~\bibnamefont{L{\"o}ffler}},
\bibinfo{author}{\bibfnamefont{C.}~\bibnamefont{Hofmann}},
\bibinfo{author}{\bibfnamefont{S.}~\bibnamefont{Kuhn}},
\bibinfo{author}{\bibfnamefont{S.}~\bibnamefont{Reitzenstein}},
\bibinfo{author}{\bibfnamefont{L.~V.} \bibnamefont{Keldysh}},
\bibinfo{author}{\bibfnamefont{V.~D.} \bibnamefont{Kulakovskii}},
\bibinfo{author}{\bibfnamefont{T.~L.} \bibnamefont{Reinecke}},
\bibnamefont{and} \bibinfo{author}{\bibfnamefont{A.}~\bibnamefont{Forchel}},
\bibinfo{journal}{Nature} \textbf{\bibinfo{volume}{432}}, \bibinfo{pages}{197
EP } (\bibinfo{year}{2004}).

\bibitem[{\citenamefont{Fischer et~al.}(2016)\citenamefont{Fischer, M{\"u}ller,
Rundquist, Sarmiento, Piggott, Kelaita, Dory, Lagoudakis, and Vu{\v
c}kovi{\'c}}}]{Fischer16}
\bibinfo{author}{\bibfnamefont{K.~A.} \bibnamefont{Fischer}},
\bibinfo{author}{\bibfnamefont{K.}~\bibnamefont{M{\"u}ller}},
\bibinfo{author}{\bibfnamefont{A.}~\bibnamefont{Rundquist}},
\bibinfo{author}{\bibfnamefont{T.}~\bibnamefont{Sarmiento}},
\bibinfo{author}{\bibfnamefont{A.~Y.} \bibnamefont{Piggott}},
\bibinfo{author}{\bibfnamefont{Y.}~\bibnamefont{Kelaita}},
\bibinfo{author}{\bibfnamefont{C.}~\bibnamefont{Dory}},
\bibinfo{author}{\bibfnamefont{K.~G.} \bibnamefont{Lagoudakis}},
\bibnamefont{and} \bibinfo{author}{\bibfnamefont{J.}~\bibnamefont{Vu{\v
c}kovi{\'c}}}, \bibinfo{journal}{Nature Photonics}
\textbf{\bibinfo{volume}{10}}, \bibinfo{pages}{163 EP }
(\bibinfo{year}{2016}).

\bibitem[{\citenamefont{Schr{\"o}der et~al.}(2017)\citenamefont{Schr{\"o}der,
Trusheim, Walsh, Li, Zheng, Schukraft, Sipahigil, Evans, Sukachev, Nguyen
et~al.}}]{Schroder17}
\bibinfo{author}{\bibfnamefont{T.}~\bibnamefont{Schr{\"o}der}},
\bibinfo{author}{\bibfnamefont{M.~E.} \bibnamefont{Trusheim}},
\bibinfo{author}{\bibfnamefont{M.}~\bibnamefont{Walsh}},
\bibinfo{author}{\bibfnamefont{L.}~\bibnamefont{Li}},
\bibinfo{author}{\bibfnamefont{J.}~\bibnamefont{Zheng}},
\bibinfo{author}{\bibfnamefont{M.}~\bibnamefont{Schukraft}},
\bibinfo{author}{\bibfnamefont{A.}~\bibnamefont{Sipahigil}},
\bibinfo{author}{\bibfnamefont{R.~E.} \bibnamefont{Evans}},
\bibinfo{author}{\bibfnamefont{D.~D.} \bibnamefont{Sukachev}},
\bibinfo{author}{\bibfnamefont{C.~T.} \bibnamefont{Nguyen}},
\bibnamefont{et~al.}, \bibinfo{journal}{Nature Communications}
\textbf{\bibinfo{volume}{8}}, \bibinfo{pages}{15376 EP }
(\bibinfo{year}{2017}).

\bibitem[{\citenamefont{Grimsmo}(2015)}]{Grimsmo15}
\bibinfo{author}{\bibfnamefont{A.~L.} \bibnamefont{Grimsmo}},
\bibinfo{journal}{Phys. Rev. Lett.} \textbf{\bibinfo{volume}{115}},
\bibinfo{pages}{060402} (\bibinfo{year}{2015}).

\bibitem[{\citenamefont{Pichler and Zoller}(2016)}]{Pichler16}
\bibinfo{author}{\bibfnamefont{H.}~\bibnamefont{Pichler}} \bibnamefont{and}
\bibinfo{author}{\bibfnamefont{P.}~\bibnamefont{Zoller}},
\bibinfo{journal}{Phys. Rev. Lett.} \textbf{\bibinfo{volume}{116}},
\bibinfo{pages}{093601} (\bibinfo{year}{2016}).

\bibitem[{\citenamefont{Gough and
James}(2009{\natexlab{a}})}]{gough_quantum_2009}
\bibinfo{author}{\bibfnamefont{J.}~\bibnamefont{Gough}} \bibnamefont{and}
\bibinfo{author}{\bibfnamefont{M.~R.} \bibnamefont{James}},
\bibinfo{journal}{Commun. Math. Phys.} \textbf{\bibinfo{volume}{287}},
\bibinfo{pages}{1109} (\bibinfo{year}{2009}{\natexlab{a}}).

\bibitem[{\citenamefont{Gough et~al.}(2017)\citenamefont{Gough, Grivopoulos,
and Petersen}}]{gough_isolated_2017}
\bibinfo{author}{\bibfnamefont{J.~E.} \bibnamefont{Gough}},
\bibinfo{author}{\bibfnamefont{S.}~\bibnamefont{Grivopoulos}},
\bibnamefont{and} \bibinfo{author}{\bibfnamefont{I.~R.}
\bibnamefont{Petersen}}, \emph{\bibinfo{title}{Isolated {{Loops}} in
{{Quantum Feedback Networks}}}}, \bibinfo{howpublished}{arXiv:1705.09916}
(\bibinfo{year}{2017}).

\bibitem[{\citenamefont{Wenner et~al.}(2014)\citenamefont{Wenner, Yin, Chen,
Barends, Chiaro, Jeffrey, Kelly, Megrant, Mutus, Neill
et~al.}}]{wenner_catching_2014}
\bibinfo{author}{\bibfnamefont{J.}~\bibnamefont{Wenner}},
\bibinfo{author}{\bibfnamefont{Y.}~\bibnamefont{Yin}},
\bibinfo{author}{\bibfnamefont{Y.}~\bibnamefont{Chen}},
\bibinfo{author}{\bibfnamefont{R.}~\bibnamefont{Barends}},
\bibinfo{author}{\bibfnamefont{B.}~\bibnamefont{Chiaro}},
\bibinfo{author}{\bibfnamefont{E.}~\bibnamefont{Jeffrey}},
\bibinfo{author}{\bibfnamefont{J.}~\bibnamefont{Kelly}},
\bibinfo{author}{\bibfnamefont{A.}~\bibnamefont{Megrant}},
\bibinfo{author}{\bibfnamefont{J.}~\bibnamefont{Mutus}},
\bibinfo{author}{\bibfnamefont{C.}~\bibnamefont{Neill}},
\bibnamefont{et~al.}, \bibinfo{journal}{Phys. Rev. Lett.}
\textbf{\bibinfo{volume}{112}}, \bibinfo{pages}{210501}
(\bibinfo{year}{2014}).

\bibitem[{\citenamefont{Srinivasan et~al.}(2014)\citenamefont{Srinivasan,
Sundaresan, Sadri, Liu, Gambetta, Yu, Girvin, and Houck}}]{Srinivasan14}
\bibinfo{author}{\bibfnamefont{S.~J.} \bibnamefont{Srinivasan}},
\bibinfo{author}{\bibfnamefont{N.~M.} \bibnamefont{Sundaresan}},
\bibinfo{author}{\bibfnamefont{D.}~\bibnamefont{Sadri}},
\bibinfo{author}{\bibfnamefont{Y.}~\bibnamefont{Liu}},
\bibinfo{author}{\bibfnamefont{J.~M.} \bibnamefont{Gambetta}},
\bibinfo{author}{\bibfnamefont{T.}~\bibnamefont{Yu}},
\bibinfo{author}{\bibfnamefont{S.~M.} \bibnamefont{Girvin}},
\bibnamefont{and} \bibinfo{author}{\bibfnamefont{A.~A.} \bibnamefont{Houck}},
\bibinfo{journal}{Phys. Rev. A} \textbf{\bibinfo{volume}{89}},
\bibinfo{pages}{033857} (\bibinfo{year}{2014}).

\bibitem[{\citenamefont{Yin et~al.}(2013)\citenamefont{Yin, Chen, Sank,
O'Malley, White, Barends, Kelly, Lucero, Mariantoni, Megrant
et~al.}}]{yin_catch_2013}
\bibinfo{author}{\bibfnamefont{Y.}~\bibnamefont{Yin}},
\bibinfo{author}{\bibfnamefont{Y.}~\bibnamefont{Chen}},
\bibinfo{author}{\bibfnamefont{D.}~\bibnamefont{Sank}},
\bibinfo{author}{\bibfnamefont{P.~J.~J.} \bibnamefont{O'Malley}},
\bibinfo{author}{\bibfnamefont{T.~C.} \bibnamefont{White}},
\bibinfo{author}{\bibfnamefont{R.}~\bibnamefont{Barends}},
\bibinfo{author}{\bibfnamefont{J.}~\bibnamefont{Kelly}},
\bibinfo{author}{\bibfnamefont{E.}~\bibnamefont{Lucero}},
\bibinfo{author}{\bibfnamefont{M.}~\bibnamefont{Mariantoni}},
\bibinfo{author}{\bibfnamefont{A.}~\bibnamefont{Megrant}},
\bibnamefont{et~al.}, \bibinfo{journal}{Phys. Rev. Lett.}
\textbf{\bibinfo{volume}{110}}, \bibinfo{pages}{107001}
(\bibinfo{year}{2013}).

\bibitem[{\citenamefont{Bader et~al.}(2013)\citenamefont{Bader, Heugel,
Chekhov, Sondermann, and Leuchs}}]{Bader13}
\bibinfo{author}{\bibfnamefont{M.}~\bibnamefont{Bader}},
\bibinfo{author}{\bibfnamefont{S.}~\bibnamefont{Heugel}},
\bibinfo{author}{\bibfnamefont{A.~L.} \bibnamefont{Chekhov}},
\bibinfo{author}{\bibfnamefont{M.}~\bibnamefont{Sondermann}},
\bibnamefont{and} \bibinfo{author}{\bibfnamefont{G.}~\bibnamefont{Leuchs}},
\bibinfo{journal}{New Journal of Physics} \textbf{\bibinfo{volume}{15}},
\bibinfo{pages}{123008} (\bibinfo{year}{2013}).

\bibitem[{\citenamefont{Korotkov}(2011)}]{korotkov_flying_2011}
\bibinfo{author}{\bibfnamefont{A.~N.} \bibnamefont{Korotkov}},
\bibinfo{journal}{Phys. Rev. B} \textbf{\bibinfo{volume}{84}},
\bibinfo{pages}{014510} (\bibinfo{year}{2011}).

\bibitem[{\citenamefont{Jahne et~al.}(2007)\citenamefont{Jahne, Yurke, and
Gavish}}]{jahne_high-fidelity_2007}
\bibinfo{author}{\bibfnamefont{K.}~\bibnamefont{Jahne}},
\bibinfo{author}{\bibfnamefont{B.}~\bibnamefont{Yurke}}, \bibnamefont{and}
\bibinfo{author}{\bibfnamefont{U.}~\bibnamefont{Gavish}},
\bibinfo{journal}{Phys. Rev. A} \textbf{\bibinfo{volume}{75}},
\bibinfo{pages}{010301} (\bibinfo{year}{2007}).

\bibitem[{\citenamefont{Cirac et~al.}(1997)\citenamefont{Cirac, Zoller, Kimble,
and Mabuchi}}]{cirac_quantum_1997}
\bibinfo{author}{\bibfnamefont{J.~I.} \bibnamefont{Cirac}},
\bibinfo{author}{\bibfnamefont{P.}~\bibnamefont{Zoller}},
\bibinfo{author}{\bibfnamefont{H.~J.} \bibnamefont{Kimble}},
\bibnamefont{and} \bibinfo{author}{\bibfnamefont{H.}~\bibnamefont{Mabuchi}},
\bibinfo{journal}{Phys. Rev. Lett.} \textbf{\bibinfo{volume}{78}},
\bibinfo{pages}{3221} (\bibinfo{year}{1997}).

\bibitem[{\citenamefont{Gough and
James}(2009{\natexlab{b}})}]{gough_series_2009}
\bibinfo{author}{\bibfnamefont{J.}~\bibnamefont{Gough}} \bibnamefont{and}
\bibinfo{author}{\bibfnamefont{M.}~\bibnamefont{James}},
\bibinfo{journal}{Autom. Control IEEE Trans. On}
\textbf{\bibinfo{volume}{54}}, \bibinfo{pages}{2530 }
(\bibinfo{year}{2009}{\natexlab{b}}).

\bibitem[{\citenamefont{Dicke}(1954)}]{dicke_coherence_1954}
\bibinfo{author}{\bibfnamefont{R.~H.} \bibnamefont{Dicke}},
\bibinfo{journal}{Phys. Rev.} \textbf{\bibinfo{volume}{93}},
\bibinfo{pages}{99} (\bibinfo{year}{1954}).

\bibitem[{\citenamefont{Axline et~al.}(2018)\citenamefont{Axline, Burkhart,
Pfaff, Zhang, Chou, Campagne-Ibarcq, Reinhold, Frunzio, Girvin, Jiang
et~al.}}]{Axline18}
\bibinfo{author}{\bibfnamefont{C.~J.} \bibnamefont{Axline}},
\bibinfo{author}{\bibfnamefont{L.~D.} \bibnamefont{Burkhart}},
\bibinfo{author}{\bibfnamefont{W.}~\bibnamefont{Pfaff}},
\bibinfo{author}{\bibfnamefont{M.}~\bibnamefont{Zhang}},
\bibinfo{author}{\bibfnamefont{K.}~\bibnamefont{Chou}},
\bibinfo{author}{\bibfnamefont{P.}~\bibnamefont{Campagne-Ibarcq}},
\bibinfo{author}{\bibfnamefont{P.}~\bibnamefont{Reinhold}},
\bibinfo{author}{\bibfnamefont{L.}~\bibnamefont{Frunzio}},
\bibinfo{author}{\bibfnamefont{S.~M.} \bibnamefont{Girvin}},
\bibinfo{author}{\bibfnamefont{L.}~\bibnamefont{Jiang}},
\bibnamefont{\textit{et~al.}}, \bibinfo{journal}{Nat. Phys.}~\bibinfo{pages}{1}
(\bibinfo{year}{2018}).

\bibitem[{\citenamefont{Hudson and Parthasarathy}(1984)}]{hudson_quantum_1984}
\bibinfo{author}{\bibfnamefont{R.~L.} \bibnamefont{Hudson}} \bibnamefont{and}
\bibinfo{author}{\bibfnamefont{K.~R.} \bibnamefont{Parthasarathy}},
\bibinfo{journal}{Commun. Math. Phys.} \textbf{\bibinfo{volume}{93}},
\bibinfo{pages}{301} (\bibinfo{year}{1984}).

\bibitem[{Note1()}]{Note1}
\bibinfo{note}{We have chosen to denote the connection matrix by
$\protect \mathbf {W}$ because in mathematical terminology it is a ``weighted
adjacency matrix''}.

\bibitem[{\citenamefont{Asenjo-Garcia et~al.}(2017)\citenamefont{Asenjo-Garcia,
Hood, Chang, and Kimble}}]{asenjo-garcia_atom-light_2017}
\bibinfo{author}{\bibfnamefont{A.}~\bibnamefont{Asenjo-Garcia}},
\bibinfo{author}{\bibfnamefont{J.~D.} \bibnamefont{Hood}},
\bibinfo{author}{\bibfnamefont{D.~E.} \bibnamefont{Chang}}, \bibnamefont{and}
\bibinfo{author}{\bibfnamefont{H.~J.} \bibnamefont{Kimble}},
\bibinfo{journal}{Phys. Rev. A} \textbf{\bibinfo{volume}{95}},
\bibinfo{pages}{033818} (\bibinfo{year}{2017}).

\bibitem[{\citenamefont{Yao et~al.}(2009)\citenamefont{Yao, Van~Vlack, Reza,
Patterson, Dignam, and Hughes}}]{yao_ultrahigh_2009}
\bibinfo{author}{\bibfnamefont{P.}~\bibnamefont{Yao}},
\bibinfo{author}{\bibfnamefont{C.}~\bibnamefont{Van~Vlack}},
\bibinfo{author}{\bibfnamefont{A.}~\bibnamefont{Reza}},
\bibinfo{author}{\bibfnamefont{M.}~\bibnamefont{Patterson}},
\bibinfo{author}{\bibfnamefont{M.~M.} \bibnamefont{Dignam}},
\bibnamefont{and} \bibinfo{author}{\bibfnamefont{S.}~\bibnamefont{Hughes}},
\bibinfo{journal}{Phys. Rev. B} \textbf{\bibinfo{volume}{80}},
\bibinfo{pages}{195106} (\bibinfo{year}{2009}).

\bibitem[{\citenamefont{Kurpiers et~al.}(2017)\citenamefont{Kurpiers, Magnard,
Walter, Royer, Pechal, Heinsoo, Salathé, Akin, Storz, Besse
et~al.}}]{Kurpiers18}
\bibinfo{author}{\bibfnamefont{P.}~\bibnamefont{Kurpiers}},
\bibinfo{author}{\bibfnamefont{P.}~\bibnamefont{Magnard}},
\bibinfo{author}{\bibfnamefont{T.}~\bibnamefont{Walter}},
\bibinfo{author}{\bibfnamefont{B.}~\bibnamefont{Royer}},
\bibinfo{author}{\bibfnamefont{M.}~\bibnamefont{Pechal}},
\bibinfo{author}{\bibfnamefont{J.}~\bibnamefont{Heinsoo}},
\bibinfo{author}{\bibfnamefont{Y.}~\bibnamefont{Salathé}},
\bibinfo{author}{\bibfnamefont{A.}~\bibnamefont{Akin}},
\bibinfo{author}{\bibfnamefont{S.}~\bibnamefont{Storz}},
\bibinfo{author}{\bibfnamefont{J.-C.} \bibnamefont{Besse}},
\bibnamefont{et~al.}, \emph{\bibinfo{title}{Deterministic quantum state
transfer and generation of remote entanglement using microwave photons}},
\bibinfo{howpublished}{arXiv:1712.08593} (\bibinfo{year}{2017}).

\end{thebibliography}

\end{document}